\documentclass[aps,pre,showpacs,reprint,superscriptaddress,amsmath,amsfonts]{revtex4-1}
\usepackage{amssymb}
\usepackage{graphicx}

\usepackage{color}

\newcommand*\kT{\ensuremath{k_\text{B}T}}
\newcommand*\Fex{\ensuremath{F_\text{ex}}}
\newcommand*\Fexc{\ensuremath{F_\text{ex}^{\text{c}}}}

\DeclareMathOperator\Real{Re}
\DeclareMathOperator\Imag{Im}
\newcommand{\A}{\tensor{A}}

\renewcommand{\tensor}[1]{\underline{\underline{#1}}}
\renewcommand{\vec}[1]{\underline{#1}}

\begin{document}

\title{Asymptotic analysis of mode-coupling theory of
  active nonlinear microrheology}
\date{\today}

\newcommand\mpi{\affiliation{Max Planck Institute for Mathematics in the Sciences, 04103 Leipzig, Germany}}
\newcommand\ukn{\affiliation{Fachbereich Physik, Universit\"at Konstanz,
  78457 Konstanz, Germany}}
\newcommand\zk{\affiliation{Zukunftskolleg, Universit\"at Konstanz,
  78457 Konstanz, Germany}}
\newcommand\dlr{\affiliation{Institut f\"ur Materialphysik im Weltraum,
  Deutsches Zentrum f\"ur Luft- und Raumfahrt (DLR), 51170 K\"oln, Germany}}
\author{M.~V.~Gnann}\mpi\ukn
\author{Th.~Voigtmann}\ukn\zk\dlr

\begin{abstract}
We discuss a schematic model of mode-coupling theory for force-driven
active nonlinear microrheology, where a single probe particle is pulled
by a constant external force through a dense host medium. The model exhibits
both a glass transition for the host, and a force-induced delocalization
transition, where an initially localized probe inside the glassy host
attains a nonvanishing steady-state velocity by locally melting the
glass. Asymptotic expressions for the transient density correlation functions of
the schematic model are derived, valid close to the transition points.
There appear several nontrivial time scales relevant for the decay laws
of the correlators. For the nonlinear friction coefficient of the probe,
the asymptotic expressions cause various regimes of power-law variation
with the external force, and two-parameter scaling laws.
\end{abstract}


\maketitle

\section{Introduction}\label{intro}

Microrheology is a modern technique that allows to probe complex
fluids on mesoscopic length scales. One inserts a probe particle,
typically $\mu$m-sized, into a host liquid of constituents that are roughly
of the same size (such as colloidal dispersions or biophysical fluids).
Monitoring the motion of the probe, one can infer
local visco-elastic response functions of the host liquid. A particularly
compelling extension of the technique is called active microrheology:
here, the probe is subjected to a controlled external drive. This is
most conveniently applied by using laser tweezers or magnetically susceptible
probe particles, or even by tailoring colloidal probes such that they
undergo self-driven motion due to chemical processes
\cite{Waigh.2005,Squires.2008,Erbe.2008}.
Active microrheology
has become a major tool in biophysics and for colloidal model systems
\cite{Wilhelm.2008,Hastings.2003,Habdas.2004,Williams.2006}.

Here we focus on force-driven active microrheology, where a constant external
force $\vec F_\text{ex}$ is applied to the probe particle. A natural quantity
to observe then is the resulting probe velocity $\vec v$, and specifically
its ensemble-averaged stationary value, $\langle\vec v\rangle_{t\to\infty}$.
The influence of the host liquid is characterized by a friction
coefficient,
\begin{equation}\label{zeta}
  \zeta\langle\vec v\rangle_{t\to\infty} = \vec{F}_{\text{ex}}\,.
\end{equation}
For typical soft-matter systems, thermal fluctuations give rise to
forces in the range of $\text{pN}$; it is hence easy to drive the system
into the nonlinear nonequilibrium regime using active microrheology.
In Eq.~\eqref{zeta}, the friction coefficient $\zeta(F_\text{ex})$ then
becomes a function of the applied force.
This makes analysis of the experiment vastly more difficult, since one
needs to employ a theory of the nonlinear probe--bath interactions.
If that is available, the technique can, however, be
rewarding, as it gives access to much more information about the complex
host liquid than a linear-response setup would. Also, compared to other
nonlinear-response techniques, such as macroscopic rheology, one gains
access to the microscopic mechanisms relevant for the dynamics
\cite{Squires.2010}.

Recently, a microscopic theory for the force-driven active nonlinear
microrheology has been proposed \cite{Gazuz.2009}, using a combination
of the integration-through transients (ITT) scheme together with
approximations inspired by the mode-coupling theory for the glass
transition (MCT).
ITT expresses the nonlinear friction coefficient
$\zeta(F_\text{ex})$ via a relation of (generalized) Green-Kubo type, through
a transient force autocorrelation function. This is a correlation function
taken with the equilibrium ensemble, but the full nonequilibrium dynamics.
Governed by the idea that in a dense system, structural relaxation via
density fluctuations is the dominant slow dynamical process,
MCT provides an approximate
closure for this correlation function in terms of (again, transient) density
correlators. These are calculated by a set of
nonlinear integro-differential equations. As a result, the nontrivial
relaxation pattern predicted for the density correlators directly gives
rise to nonlinearities in the friction coefficient.

So far, the full MCT-ITT equations for active microrheology have proven
not amenable to extended numerical treatment. The strategy used in
Ref.~\cite{Gazuz.2009} therefore was to implement an ad-hoc simplified
model that reduces the complexity down to a single fluctuation mode
in the hope of retaining all nontrivial mathematical features of the
original set of equations. Introducing a few adjustable parameters to this
so-called schematic model, a successful quantitative analysis of available
computer-simulation and experimental data provides an \textit{ex posteriori}
justification for doing so. The original schematic model proved too
restrictive in certain aspects. Recently, an improved
version of the model has been presented \cite{Gnann.2011}, which allows
convincing fits of the data in all accessible regimes by taking into
account some aspects of the force-induced spatial anisotropy in the dynamics.

The data show a strong nonlinearity in $\zeta(F_\text{ex})$: in a
relatively narrow range of forces, close to the glass transition
the friction coefficient drops by
orders of magnitude, separating a near-equilibrium low-force regime from
a high-force regime. The schematic MCT models interpret the resulting strong
drop as a precursor to a probe-delocalization transition: inside a glassy
host, the probe particle
is held in a nearest-neighbor cage (formally $\zeta\to\infty$ at low
forces, for the ideal-glass case) that can sustain a finite amount of
external force. Once the applied force exceeds a critical threshold
$\Fexc$, the probe's nearest-neighbor cage is forced open so that
a finite mean velocity results (and $\zeta$ drops to a finite value).
The threshold force is thus interpreted as a measure of the cage strength.
Since cages are formed in a highly collective process involving the host
particles and their local structure, active nonlinear microrheology is
in principle a unique tool to probe the local rigidity of that structure.
In fact, $\Fexc=\mathcal O(50\kT/\sigma)$ has been measured
\cite{Gazuz.2009}, greatly exceeding the typical force scale of thermal
fluctuations, $\kT/\sigma$ (here, $\sigma$ is a typical host-particle
size).

In order to understand the asymptotic behavior of $\zeta(F_\text{ex})$
close to the critical force within ITT-MCT,
the asymptotic behavior of the correlation
functions close to this delocalization transition has to be understood.
In essence, a two-parameter scaling prescription is sought for, since one
is dealing with both the distance to the glass transition, and the distance
to the delocalization transition, as small parameters.

In this paper, we present an asymptotic treatment of the schematic
ITT-MCT equations for force-driven active nonlinear rheology. We discuss
the set of equations that has been solved numerically in
Ref.~\cite{Gnann.2011}, where the model was shown to describe both computer
simulation and experiment quantitatively.
The calculation proceeds by considering the double limit of approaching
the glass transition and the delocalization transition, which yields two
small parameters and various asymptotic results depending on their ratio
and sign. The techniques we use are similar to those that have been used
earlier to derive different two-parameter scaling laws within MCT,
e.g., for the extended mode-coupling theory including a schematic
hopping term \cite{Goetze.1987}, or for the ITT-MCT equations describing
the macrorheology for a given constant shear rate \cite{Fuchs.2003}.

Our analysis, however, has aspects that differ from the above-mentioned cases.
Due to the nonequilibrium nature of the
problem, the time-evolution operator is non-Hermitian, and gives rise
to complex-valued correlation functions. Usually, in the regime of
structural relaxation, one can safely assume the correlation functions
appearing in the theory to be real-valued and completely monotone; this
holds rigorously for overdamped short-time dynamics as applicable
to colloidal suspensions in equilibrium
\cite{Goetze.1995,Franosch.2002}. These properties are used in the
asymptotic expansion, for example to ensure that the singularities to be
discussed belong to a certain class of bifurcations \cite{Goetze.2009}.
Even in the macro-rheology of colloidal suspensions, where external flow is
represented by a non-Hermitian generalized Smoluchowski operator,
taking into account the mechanism of shear advection separately allows one
to return to real-valued correlation functions and to a scheme of asymptotic
expansions that closely mirrors the one followed in equilibrium
\cite{Fuchs.2009,Hajnal.2009,Krueger.2011}.

In the equations for active microrheology, some assumptions entering the
standard discussion of MCT are no longer valid. Consequently, the mathematical
classification of the transition between localized and delocalized probe
particles (in the idealized glass) is still open. We restrict ourselves
here to a certain schematic model that is inspired by, but not necessarily
mathematically equivalent to, the microscopic equations presented in
Ref.~\cite{Gazuz.2009}. Since the model has been successfully used in
data analysis, the restriction appears plausible.

The paper is organized as follows: in Sec.~\ref{mct}, we summarize the
equations defining the model. The long-time limits of the correlation
functions, characterizing the glassy and localized states, are discussed
in Sec.~\ref{sec:longtime}. Sections~\ref{beta} and \ref{alpha} are devoted
to deriving asymptotic expressions for times large compared to the
single-particle relaxation time, valid on intermediate- respectively
long-time windows that open upon approaching the transition points. These
are the analogues to the common MCT scaling laws referred to as
$\beta$- and $\alpha$-scaling. In Sec.~\ref{sec:zeta} we transfer these
results to a two-parameter scaling law for the friction coefficient,
after which Sec.~\ref{conclusion} concludes.

\section{Schematic Mode-Coupling Theory}\label{mct}

We summarize the main equations defining the schematic MCT model
for active nonlinear microrheology.
Following a generic integration-through-transients (ITT) scheme
and the notion that the slow dynamics in the vicinity of the (colloidal)
glass transition is dominated by density fluctuations
\cite{Fuchs.2002}, the central quantities of the model are the
transient density correlation functions. While in the liquid state, these
eventually decay to zero, in the glass they attain a finite positive long-time
limit called the glass form factor or nonergodicity parameter $f$.
Even in the liquid,
the correlation functions stay close to its value $f_c$ at the transition
over an increasingly large time window as one approaches the transition.
The quantity $f$ and its tagged-particle counterpart $f^s$ will play a
central role in the discussion to follow.

In the specific model we choose,
there is one correlator $\phi(t)$ mimicking the dynamics of the host
liquid and determining the glass form factor $f=\lim_{t\to\infty}\phi(t)$.
Furthermore, this dynamics is taken as identical to the equilibrium
one. Clearly, in the thermodynamic limit and assuming that the probe--host
interactions remain sufficiently short-ranged, the host-liquid dynamics
is in the ensemble average unperturbed by the external force that is applied
to the probe particle only. The macroscopic state of the host (liquid or
glassy) will then be determined by the equilibrium coupling coefficients
only.
The equation of motion for the host-liquid correlator takes the form
\begin{subequations}\label{f12}
\begin{equation}\label{f12phi}
  \partial_t\phi(t)+\Gamma\left\{\phi(t)+\int_0^tm(t-t')
  \partial_{t'}\phi(t')\,dt'\right\}=0\,.
\end{equation}
Equations of this type can be derived from microscopic starting points using
a Mori-Zwanzig projection operator scheme \cite{Goetze.2009}, where the
slow relaxation is modeled by a memory kernel $m(t)$ determined by the
fluctuating forces. The coefficient
$\Gamma$ is a relaxation rate of the short-time dynamics, and will be
set to unity in the calculations below (thus defining the unit of time).
The central idea of MCT is to approximate the memory kernel by a bilinear
form of the correlators themselves, expressing the notion that a slow
decay of density fluctuations leads to and hinges upon slow decorrelation
of fluctuating forces. In our discussion we adopt the so-called
$\text{F}_\text{12}$ model,
\begin{equation}\label{f12m}
  m(t)=v_1\phi(t)+v_2\phi(t)^2\,,
\end{equation}
with positive parameters $(v_1,v_2)$.
\end{subequations}
With this choice, the full range of asymptotic behavior expected from the
microscopic MCT close to ordinary liquid--glass transitions is reproduced
\cite{Goetze.2009}.
The use of one single mode to describe the
host dynamics embodies our assumption that the system remains homogeneous
and isotropic in the ensemble average.

Spatial isotropy clearly breaks down for the motion of the probe particle.
Therefore, in the schematic model of Ref.~\cite{Gnann.2011}, two tagged-particle
correlation functions $\phi^s_\alpha(t)$ were introduced, labeled by
$\alpha\in\{\parallel,\perp\}$ to indicate the separate role played by
fluctuations in direction of, and perpendicular to, the applied force.
The equations of motion then read
\begin{subequations}\label{sjoe}
\begin{equation}\label{sjoephi}
  \partial_t\phi^s_\alpha(t)+\omega_\alpha\left\{
  \phi^s_\alpha(t)+\int_0^tm^s_\alpha(t-t')\partial_{t'}\phi^s_\alpha(t')\,dt'
  \right\}=0\,.
\end{equation}
Here, the initial relaxation rates are $\omega_\perp=\Gamma_s$ and
$\omega_\parallel=\Gamma_s(1-i\kappa_\parallel\Fex)$, which fixes the unit of energy
as $\kT=1$.
We furthermore set $\Gamma_s=1$ for simplicity, suitable for a probe
that is almost identical to the host particles.
$\kappa_\parallel$ adjusts the scale of forces entering the schematic model.
The slow dynamics of the probe arises as a consequence of slow dynamics in
the host liquid, and is modeled by memory kernels containing a linear
coupling to $\phi(t)$,
\begin{gather}\label{sjoemparallel}
  m^s_\parallel(t)=\left(v_1^s\phi^{s*}_\parallel(t)
  +v_2^s\phi^s_\perp(t)\right)\phi(t)/(1-i\kappa_\parallel\Fex)\,,\\
  \label{sjoemperp}
  m^s_\perp(t)=\left(v_1^s\phi^s_\perp(t)+v_2^s\Real\phi^s_\parallel(t)
  \right)\phi(t)/(1+(\kappa_\perp\Fex)^2)\,.
\end{gather}
\end{subequations}
The specific choice of terms entering Eqs.~\eqref{sjoe} is rooted in
symmetry considerations based on the full microscopic MCT model
\cite{Gazuz.2009,Gnann.2011}.
Specifically, for $\Fex=0$, the model reduces to a well-studied schematic
model of tagged-particle motion close to the glass transition, the
Sj\"ogren model \cite{Sjoegren.1986}. In this case $\phi^s_\parallel(t)
\equiv\phi^s_\perp(t)$, both are real-valued, and only a single coupling
strength $v^s=v_1^s+v_2^s$ is relevant.
Assuming that the probe motion retains
rotational symmetry around the axis set by $\Fex$ in the ensemble average,
it is seen that $\phi^s_\perp(t)$ remains a real-valued function.
On the other hand, a nonvanishing net displacement of the probe along the
direction of the force is expected, which results in the modulation of
the corresponding Fourier-transformed density-fluctuation
correlations with a complex phase. Hence,
$\phi^s_\parallel(t)$ will be complex-valued. Furthermore,
Eqs.~\eqref{sjoe} obey the expected symmetries under
reversal of the applied force, $\Fex\mapsto-\Fex$. The parameters
$\kappa_\parallel$ and $\kappa_\perp$ characterize the
forces relevant for inducing the decay of fluctuations in the
two directions. They are introduced to achieve
quantitative fits of simulation and experimental data \cite{Gnann.2011}.
We fix $\kappa_\perp=1/2$ and $\kappa_\parallel=1$
for the numerical calculations presented below.

We are thus left with a three-correlator model where four parameters
enter that are relevant for an understanding of the qualitative long-time
behavior: two of them, $(v_1,v_2)$ model the approach to the glass
transition in the host liquid. In the parameter space of the
$\text{F}_\text{12}$ model, a line of ideal glass transitions $(v_{1,c},v_{2,c})$
exists, and it is useful to characterize the distance to any chosen point
on this line by a single distance parameter $\varepsilon$. For the
discussion below, let us fix $v_{2,c}=2$ implying $v_{1,c}=2(\sqrt2-1)$
\cite{Goetze.2009}, and set
$(v_1,v_2)=(v_{1,c},v_{2,c})(1+\varepsilon)$. As usual, $\varepsilon<0$ signals
liquid states, while $\varepsilon>0$ holds for glassy states.
We will denote quantities calculated at glass-transition points of the
host-liquid model by subscripts $c$.

The other two parameters, $(v_1^s,v_2^s)$, represent the coupling of
the probe to the host liquid; they will, among other things, also reflect
a non-trivial size ratio between probe and host particles.
We employ the simplification
that was used in Ref.~\cite{Gnann.2011} and set $v_1^s/v_2^s=2$ for
numerical calculations.
Note that setting $v_2^s=0$ in Eqs.~\eqref{sjoe} reduces the model to the
one originally proposed in Ref.~\cite{Gazuz.2009}, not taking into account
the role of probe-density fluctuations perpendicular to the force direction.

While the transient correlation functions can in principle be measured
and have been evaluated in computer simulation \cite{Gazuz.2009,Gnann.2011},
an experimentally more easily accessible quantity is the friction coefficient
$\zeta$. A straightforward adaption of the microscopic expression,
used in \cite{Gnann.2011}, is $\zeta=1+\Delta\zeta$ with
\begin{multline}\label{delzeta}
\Delta \zeta  = \mu\int_0^\infty \phi(t)\phi^s_\perp(t)\,dt\\
  +(1-\mu)\int_0^\infty \phi(t)\Real\phi^s_\parallel(t)\,dt\,.
\end{multline}
This uses that in our choice of units, the solvent friction experienced
by the free particle is unity, $\zeta=1$.
In the microscopic theory, $\Delta\zeta$ is given as an angular average
over a force-autocorrelation function; approximating the latter in terms
of density-pair modes, one gets a Fourier-space integral over anisotropic
coupling coefficients.
In writing Eq.~\eqref{delzeta}, we assume that this integral
is qualitatively dominated by contributions from the two modes considered
in the schematic model, and that in particular the qualitative features
of the correlators $\alpha=\parallel$ and $\alpha=\perp$ are not restricted
to zero-measure portions of wave-vector space.
The parameter $\mu$ allows to reweight these
contributions, which does not qualitatively change the features close
to the delocalization transition. Following Ref.~\cite{Gnann.2011}
we set $\mu=1/2$ in our calculations.

\begin{figure}
\centerline{\includegraphics[width = \columnwidth]{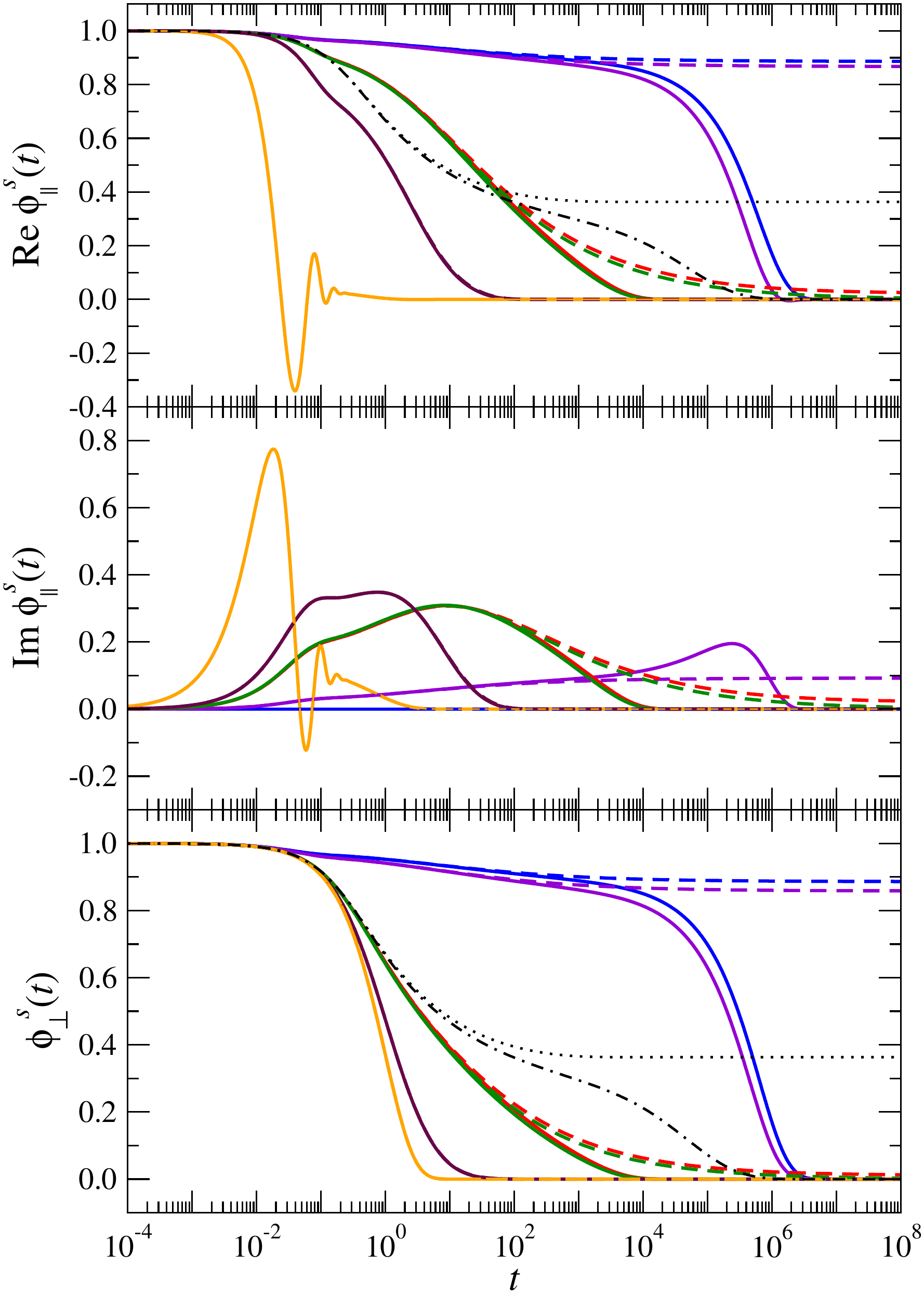}}
\caption{\label{fig:corr_overview}
  Schematic-model correlation functions $\phi^s_\parallel(t)$ and
  $\phi^s_\perp(t)$ for host-liquid parameters close to and at the glass
  transition, $\varepsilon=-10^{-2}$ (solid lines) and $\varepsilon=0$
  (dashed), for probe-coupling coefficients $v^s=30$, $\kappa_\perp=0.5$,
  and $\kappa_\parallel=1$.
  For $\phi^s_\parallel(t)$, both real and imaginary part are shown in
  separate panels.
  Curves in the order of decreasing relaxation time correspond to
  $\Fex=0$, $1$, $\Fexc-0.1$, $\Fexc$, $12$, and $80$, where $\Fexc = 6.5735$.
  In the panels showing the real parts,
  the host-liquid correlation functions $\phi(t)$ are added for the
  liquid (dash-dotted lines) and glassy (dotted) state.
}
\end{figure}

Figure~\ref{fig:corr_overview} displays exemplary correlation functions of the
schematic model both in the liquid state (solid lines) and at the
liquid--glass transition point (dashed).
Numerical solutions of Eqs.~\eqref{f12} to \eqref{sjoe} are obtained by
integrating in the
time domain, using a repeated doubling of the integration step to allow
covering a large number of decades in time. The algorithm is a straightforward
generalization of the one used in previous MCT calculations \cite{Gnann.2009}.

Considering first $\Fex=0$,
the correlation functions are all real, and Fig.~\ref{fig:corr_overview}
demonstrates
the two-step decay typical for glassy structural relaxation: at times
much larger than those associated with single-particle motion,
$t\gg1/\Gamma$, a window of structural relaxation opens. Correlators
first decay towards a finite plateau, identifying by
$\phi^s_\alpha(t\approx t_\sigma)\approx f^s_\alpha$ the so-called
$\beta$-relaxation regime. The time scale $t_\sigma$ diverges approaching
the glass transition. In the glass, and right at the glass transition
($\varepsilon=0$), the correlation functions never decay from their
plateau. In the liquid, a final decay to zero sets in on time scales
large compared to those of the $\beta$ relaxation, $t\gg t_\sigma$. This
identifies the $\alpha$-relaxation window $t/t_\sigma'=\mathcal O(1)$ and a
second time scale $t_\sigma'$ that diverges faster than $t_\sigma$ upon
approaching the glass transition.
The equilibrium
tagged-particle correlators inherit these properties from the host-liquid
correlator $\phi(t)$, shown in
Fig.~\ref{fig:corr_overview} as dash-dotted (liquid) and dotted (glass) lines.
Linear response is the regime for small $\Fex$ where the $\phi^s_\alpha(t)$
are still close to their $\Fex=0$ limiting cases, as in this case, $\zeta$
defined through Eq.~\eqref{delzeta} remains force-independent.

Increasing $\Fex$ in the schematic model leads to a decrease of
the plateau in the $\phi^s_\alpha(t)$, and to a decrease
of their $\alpha$-relaxation time. It can be argued \cite{Gazuz.2009} that
this corresponds to the fact that the localized probability density for
the location of the probe particle continuously broadens when increasing
$\Fex$. For large enough external forces, the plateau vanishes completely,
corresponding to delocalized probe motion. This allows to define a
critical or threshold force $\Fexc$. For glassy states, this indicates
the force needed to locally melt the glass surrounding the probe.
In the liquid, the nearest-neighbor cages giving rise to glassy dynamics
still persist over a time scale $t_\sigma'$, and around $\Fexc$ these are
broken faster and more effectively by the applied force than by thermal
fluctuations.
For the parameters used in Fig.~\ref{fig:corr_overview}, the following
analysis confirms $\Fexc\approx6.5735$.

At still larger forces, the correlator $\phi^s_\parallel(t)$ in
Fig.~\ref{fig:corr_overview} shows oscillatory behavior.
This can be connected to a finite average probe motion in the delocalized
state, as we will investigate in more detail below.
For $\phi^s_\perp(t)$, no such oscillations are seen, as on average
the probe will not move perpendicular to the direction of the applied
force. The oscillations are a clear signature of the non-equilibrium
nature of the dynamics, since for colloidal dynamics, the negative
semidefiniteness of the time-evolution operator rules them out
in equilibrium
\cite{Naegele.1996,Goetze.1995,Franosch.2002}.

\begin{figure}
\centerline{\includegraphics[width = \columnwidth]{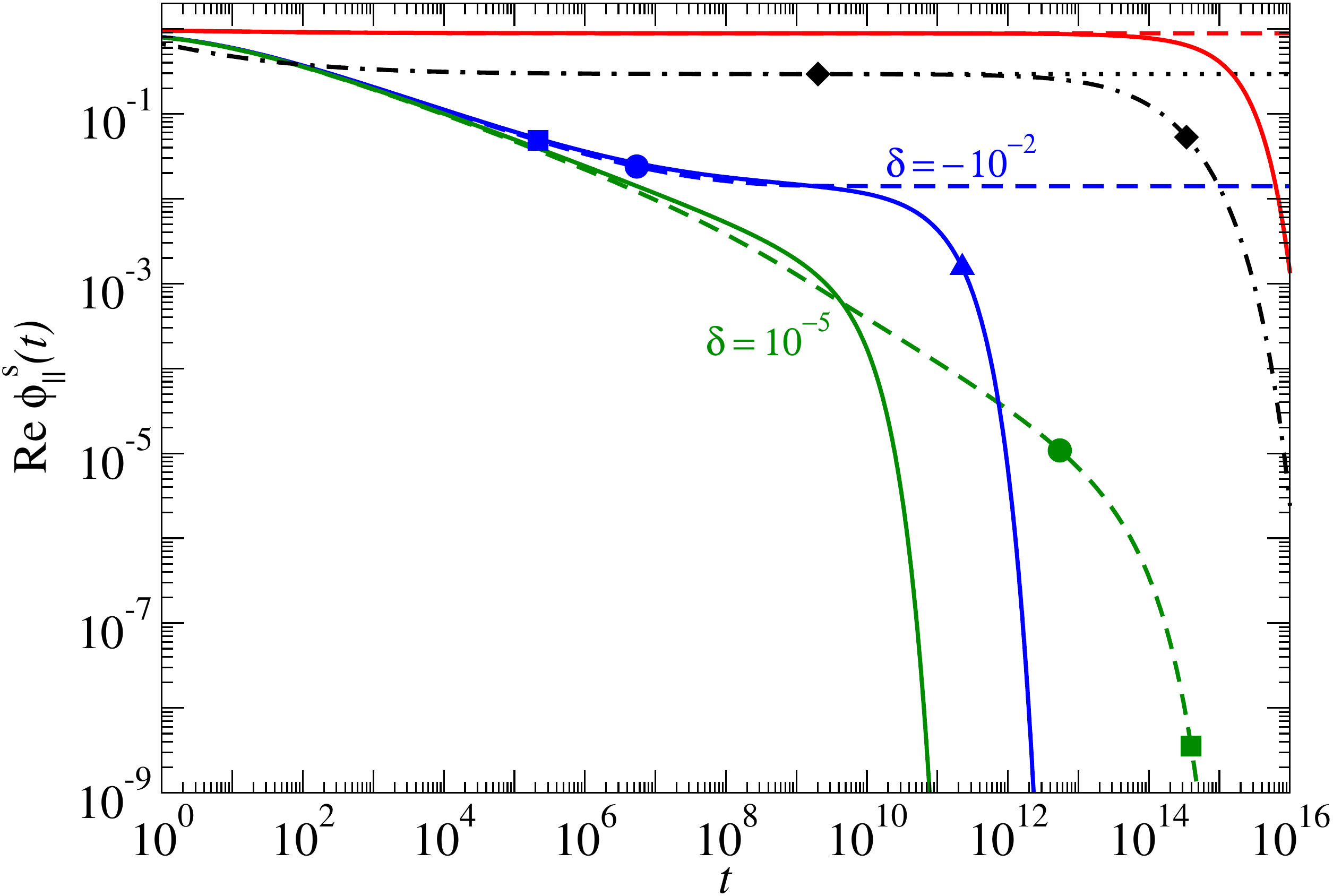}}
\caption{\label{fig:corr_overview_2}
  Real part of the probe-particle correlation functions
  $\Real\phi_\parallel^s(t)$ for the schematic
  model for force-driven microrheology, for a distance to the host-liquid
  glass transition $\varepsilon=\pm10^{-6}$, and forces
  $\Fex=\Fexc(1+\delta)$ with $\Fexc\approx6.587$ and
  $\delta=-1$, $-0.01$, and $10^{-5}$, from top to bottom.
  Model parameters are $v^s=30$, $\kappa_\perp = 0.5$, and $\kappa_\parallel = 1$ (others as mentioned in
  the text).
  Solid (dashed) lines show the results for the liquid (glass).
  Dash-dotted (dotted) lines show the corresponding host-liquid correlator
  $\phi(t)$ for the liquid (glass).
  Symbols mark the time scales
  $t_\sigma$ and $t_\sigma'$ for the host correlator (diamonds, cf.\
  Eq.~\eqref{bathtimescales}),
  $t_\delta$ (squares), $t_{1/2}$ (circles), and $t_{\sigma,\delta}'$
  (triangles) for the probe correlator (Eqs.~\eqref{tdelta}, \eqref{t12},
  and \eqref{tsigdelta}).
}
\end{figure}

Figure~\ref{fig:corr_overview_2} presents curves for the real part of
the probe-particle correlation function corresponding to fluctuations
in the direction of the force, $\Real\phi^s_\parallel(t)$, in a
double-logarithmic plot. Compared to
Fig.~\ref{fig:corr_overview}, a state even closer to the glass transition
has been taken, in order to bring out more clearly the different time
scales and the associated relaxation laws for the correlators.
To explain these is the aim of the discussion pursued in this paper.
An interesting feature brought out by the analysis presented
below is the fact that all the probe-particle correlation functions are,
asymptotically close to the glass- and the delocalization transition,
proportional to each other. Hence, a discussion of the time scales seen
in $\Real\phi^s_\parallel(t)$ suffices. They are marked in
Fig.~\ref{fig:corr_overview_2} by various symbols, and we will come back
to their discussion below.

\begin{figure}
\centerline{\includegraphics[width = \columnwidth]{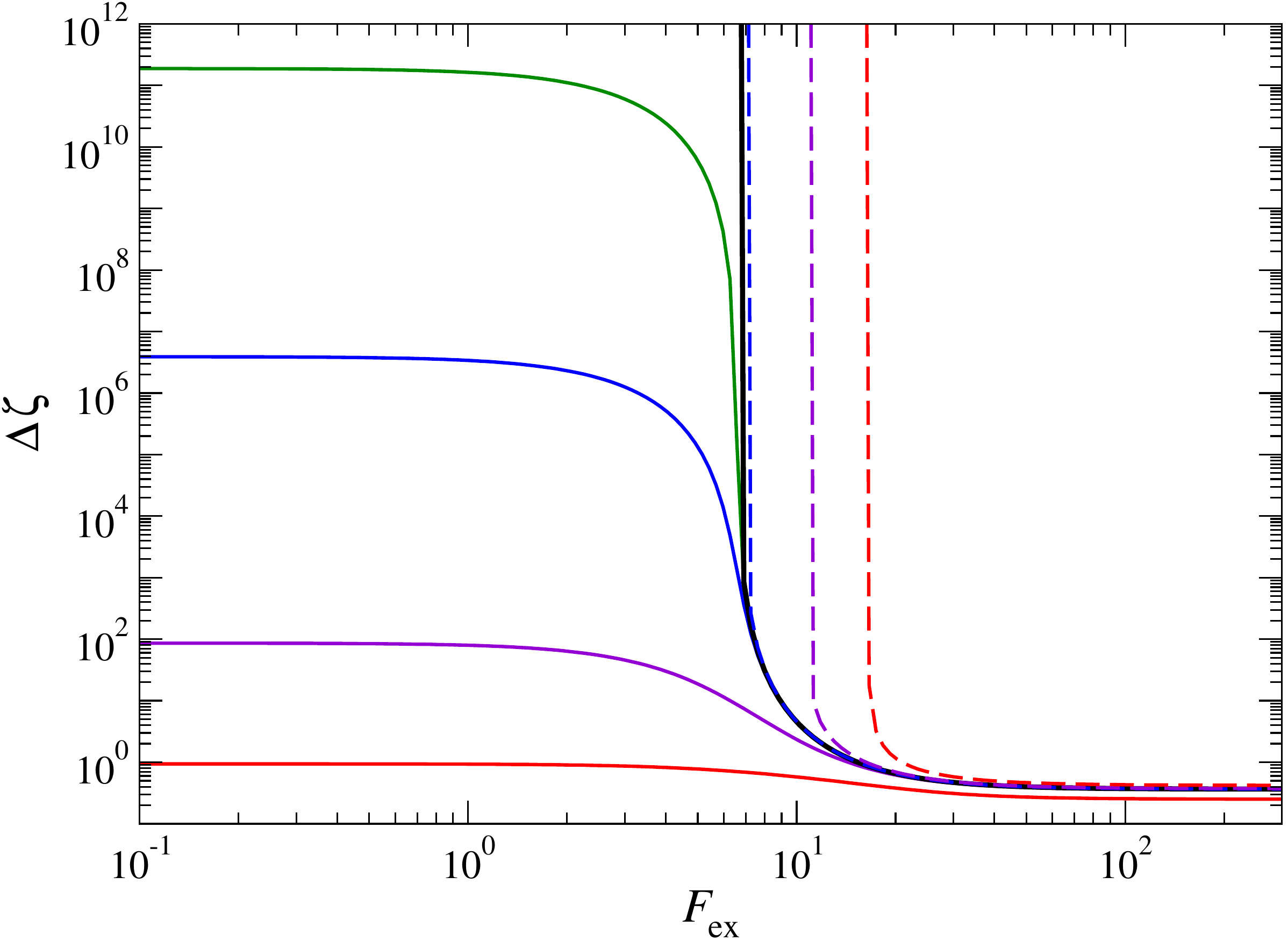}}
\caption{\label{fig:zeta_overview}
  Excess friction $\Delta\zeta$ experienced by a probe particle, as
  a function of the applied force $\Fex$, calculated in the schematic MCT
  model according to Eq.~\eqref{delzeta}. Curves from bottom to top correspond
  to distances to the host-liquid glass transition $\varepsilon=-1$,
  $-0.1$, $-10^{-3}$, $-10^{-5}$ (liquid; solid lines), $0$ (solid),
  $10^{-3}$, $0.1$, and $1$ (glass; dashed lines).
  Other parameters are chosen as in Fig.~\ref{fig:corr_overview}.
}
\end{figure}

The behavior of the correlation functions shown in
Figs.~\ref{fig:corr_overview} and \ref{fig:corr_overview_2}
gives rise to a strongly nonlinear signature in the friction coefficient.
Figure~\ref{fig:zeta_overview} shows the friction increment
$\Delta\zeta$ calculated by Eq.~\eqref{delzeta} as a function of external
force $\Fex$ for several values of $\varepsilon$; other parameters were
chosen as in Fig.~\ref{fig:corr_overview}. Qualitatively,
the resulting $\zeta(\Fex)$ agree with those discussed in
Refs.~\cite{Gazuz.2009,Gnann.2011} in conjunction with experimental and
simulation data.
For $\Fex\to0$, a linear-response regime is recovered
where $\zeta$ depends only weakly on $\Fex$. $\zeta(\Fex\to0)$ increases
strongly with decreasing $|\varepsilon|$ from the liquid side. This is
the manifestation of the glass transition, where the equilibrium mobility
of the tracer particle vanishes, as long as the coupling between probe and
host liquid is sufficiently strong. The possibility of a decoupling of
tracer motion from the host even in equilibrium (realized, e.g., by tracers
of sufficiently small size in binary mixtures of hard spheres
\cite{Voigtmann.2011b}) will not be discussed in this paper.

As $\Fex$ is increased, deviations from linear response set in quadratically
with the force, $\Delta\zeta\approx\Delta\zeta(\Fex\to0)-a\Fex^2$,
with a prefactor $a$ depending on $\varepsilon$.
A steep descent of $\Delta\zeta$ is then seen in Fig.~\ref{fig:zeta_overview}
around the threshold force,
$\Fex\approx\Fexc$.
In the glass, the point $\Fex=\Fexc$ marks the divergence of $\Delta\zeta$ as one
approaches the critical force from above, $\Fex\to\Fexc+0$.
At forces much larger than $\Fexc$, a second plateau is observed in
$\Delta\zeta$. Intuitive reasoning might suggest
$\zeta(\Fex\to\infty)\to1$ (the solvent friction in our units),
hence $\Delta\zeta=0$ in this window.
This is not seen in experiment or simulation. It was argued in
Ref.~\cite{Gnann.2011} that within ITT-MCT, the fact that
$\Delta\zeta>0$ for $\Fex\to\infty$ can be understood
by accounting for contributions to $\Delta\zeta$ stemming from
$\alpha=\perp$, i.e., fluctuations in the direction perpendicular to the
force. Note that no reference to a suspending liquid is made in this
argument, although hydrodynamic interactions mediated through the solvent
are expected to greatly influence the friction coefficient for large
forces in real colloidal suspensions.
The ratio of the large-force plateau to the linear-response value has been
determined in a low-density expansion by Brady and coworkers
\cite{Squires.2005,Carpen.2005} to be
$\Delta\zeta(\Fex\to\infty)/\Delta\zeta(\Fex\to0)=2$; this is confirmed
by simulations of dilute host liquids.
In the schematic model, the parameter $\mu$ serves to
reproduce this ratio when
one assumes the coupling coefficients $(v_1,v_2)$ and $(v_1^s,v_2^s)$
to approach zero at low densities: dropping all memory kernels from
Eqs.~\eqref{f12} and \eqref{sjoe}, the correlation functions become
$\phi(t)=\phi_\perp(t)=\exp(-t)$ and $\Real\phi_\parallel(t)=\exp(-t)
\cos(\Fex t)$. In Eq.~\eqref{delzeta} this yields
$\Delta\zeta_0=(4+\mu\Fex^2)/(8+2\Fex^2)$, which
interpolates between $\Delta\zeta_0(\Fex\to0)=1/2$ and
$\Delta\zeta_0(\Fex\to\infty)=\mu/2$. In Fig.~\ref{fig:zeta_overview}
this is not seen, since we kept $v^s=30$ fixed. While fits to data
and the microscopic theory
suggest to change $v^s$ as a function of density \cite{Gnann.2011},
for a discussion of the features close to the glass transition density
we assume $v^s\approx v^s(\varepsilon=0)$ for simplicity.

We now focus on the analytic discussion of the above observed crossovers.
For this purpose, let us employ a more compact notation of
Eq.~\eqref{sjoe}. We introduce a vector of probe correlation
functions, $\vec\phi^s(t)$, with components $(\phi^s_1(t),\phi^s_2(t),
\phi^s_3(t))^\top:=(\Real\phi_\parallel^s(t),\Imag\phi_\parallel^s(t),
\phi_\perp^s(t))^\top$, and initial conditions
$\vec\phi^s_0=(1,0,1)^\top$. We will refer to the components of this
vector with Latin indices $j\in\{1,2,3\}$ (recall that Greek indices,
$\alpha\in\{\parallel,\perp\}$, label the directions). Let us further introduce
a matrix characterizing the short-time motion,
\begin{subequations}
\begin{equation}
  \tensor\omega:=\begin{pmatrix}1&\kappa_\parallel\Fex &0\\
  -\kappa_\parallel\Fex&1&0\\ 0&0&1\end{pmatrix}
  \,,
\end{equation}
and a matrix constructing the memory kernel vector,
\begin{equation}
  \tensor{\mathcal M}^s:=
  \begin{pmatrix}\xi_\parallel\\ &\xi_\parallel\\ &&\xi_\perp\end{pmatrix}^{-1}
  \begin{pmatrix}v_1^s & \kappa_\parallel\Fex v_1^s
    & v_2^s\\
    \kappa_\parallel\Fex v_1^s & -v_1^s
    & \kappa_\parallel\Fex v_2^s\\
    v_2^s & 0 & v_1^s \end{pmatrix}
\end{equation}
with $1/\xi_\alpha=1+(\kappa_\alpha\Fex)^2$.
The complex-number multiplication appearing in the convolution integral
is then represented in our matrix notation
by a bilinear (but nonsymmetric) mapping
$\vec{\mathcal C}_j^s[\vec x,\vec y]:=\vec x^\top\cdot{\tensor{\mathcal M}^s}^\top\tensor C_j\cdot\vec y$,
with
\begin{align*}
  \tensor C_1&=\left(\begin{smallmatrix} 1\\ &-1\\ &&0\end{smallmatrix}\right)
  \,, &
  \tensor C_2&=\left(\begin{smallmatrix} &1\\ 1\\ &&0\end{smallmatrix}\right)
  \,, &
  \tensor C_3&=\left(\begin{smallmatrix} 0\\ &0\\ &&1\end{smallmatrix}\right)
  \,.
\end{align*}
It will occasionally be useful to introduce the symmetrized version of this
mapping, $\vec{\mathcal D}^s[\vec x,\vec y]=
\vec{\mathcal C}^s[\vec x,\vec y]+\vec{\mathcal C}^s[\vec y,\vec x]$.
\end{subequations}
Applying the Laplace transform,
$
\hat{\phi}(z) = i \int_0^{\infty} e^{i z t} \phi(t) dt
$,
the equations of motion Eqs.~\eqref{f12phi} and \eqref{sjoephi}
are rewritten as
\begin{subequations}\label{equmotlaplace}
\begin{equation}\label{f12philt}
z \hat{\phi}(z) - (1 + z \hat{\phi}(z)) (i z + z \hat{m}(z)) = 0
\end{equation}
for the bath correlator, and
\begin{multline}\label{sjoephilt}
-iz\tensor{\omega}^{-1}
\left(z \hat{\vec{\phi}}^s(z) + \vec\phi_0^s\right) + z \hat{\vec{\phi}}^s(z)\\
- z\vec{\mathcal C}^s\left[\widehat{\phi\vec{\phi}^s}(z), z \hat{\vec{\phi}}^s(z) + \vec\phi_0^s\right] = 0
\end{multline}
for the probe correlator.
\end{subequations}

\section{Long-Time Behavior}\label{sec:longtime}

We assume the long-time limits of all correlation functions of the
schematic model to exist. In view of standard features of structural
relaxation dynamics close to a glass transition, this is not unreasonable.
Then, the Abelian theorem \cite{Widder.1972},
\begin{subequations}\label{longtime}
\begin{gather}
f := \lim_{t \to \infty} \phi(t) = - \lim_{z \to + i 0} z \hat{\phi}(z)\,,\\
\vec{f}^s := \lim_{t \to \infty} \vec{\phi}^s(t) = - \lim_{z \to + i 0} z \hat{\vec{\phi}}^s(z)\,,
\end{gather}
\end{subequations}
leads from Eqs.~\eqref{equmotlaplace} to
\begin{equation}\label{longtimebath}
 f/(1 - f) = v_1 f + v_2 f^2
\end{equation}
for the bath correlator and
\begin{gather}
0 = \tensor{A}^s \cdot \vec{f}^s - f \vec{\mathcal{C}}^s[\vec{f}^s, \vec{f}^s]\,,\label{longtimetracer}
\end{gather}
for the probe correlators. Here we have defined
\begin{subequations}\label{longtimetracerA}
\begin{equation}
\tensor{A}^s \cdot \vec{f}^s := f \vec{\mathcal{C}}^s[\vec{f}^s, \vec\phi_0^s] - \vec{f}^s\,.\end{equation}
Explicit evaluation leads to
\begin{equation}
  \tensor{A}^s = \begin{pmatrix}
    \xi_\parallel v_1^sf-1 & \xi_\parallel\kappa_\parallel\Fex v_1^s f
    & \xi_\parallel v_2^s f\\
    \xi_\parallel\kappa_\parallel\Fex v_1^sf & -\xi_\parallel v_1^sf-1
    & \xi_\parallel\kappa_\parallel\Fex v_2^sf\\
    \xi_\perp v_2^sf &0 & \xi_\perp v_1^sf-1
  \end{pmatrix}\,.
\end{equation}
\end{subequations}

Equation~\eqref{longtimebath} describes the well-known bifurcation scenario of
the glass transition within MCT \cite{Goetze.2009}: it is an implicit
nonlinear equation for the nonergodicity parameter $f$.
Possibly many different solutions of this equation exist for general models,
with zero being an obvious one. It can be proven
\cite{Goetze.1995} that within equilibrium MCT, $f$ is always determined
by the largest positive real solution. Since for $v_{1,2}\to0$, only
$f=0$ survives, and for $v_{1,2}\to\infty$, $f>0$ holds, the MCT solutions
changes at some bifurcation point. These points define a hypersurface
$(v_{1,c},v_{2,c})$ through
$
  v_{1,c}=2\sqrt{v_{2,c}}-v_{2,c}
$ (restricting to $1<v_{2,c}\le4$).
The bifurcation is identified as the ideal glass
transition. Generically, in Eq.~\eqref{longtimebath}, a jump in $f$
occurs; this is called a type~B transition in the literature.
In other words, $f_c>0$ holds for the critical nonergodicity parameter
evaluated right at the transition.
Mathematically, one deals for generic mode-coupling models
with bifurcations of the class $\mathcal A_\ell$ \cite{Goetze.1995}
according to the Arnol'd classification \cite{Arnold.1992}; in the present
case we are only concerned with the nondegenerate
$\mathcal A_2$ bifurcations displayed by the $\text{F}_\text{12}$ model.
The bifurcation points are identified by recognizing that, as two
branches of solutions of the implicit equation for $f$ coalesce, the
equation is no longer invertible (and the implicit-function theorem violated).

Equation~\eqref{longtimetracer} describes localization respectively
delocalization of probe
particles: if $f=0$, also $\vec f^s=0$, i.e.\ in a liquid, all probe
particles are delocalized (and able to undergo long-range motion, as required
for a liquid). If $f>0$, sufficiently large
$(v_1^s,v_2^s)$ will for $\Fex=0$ lead to $f^s_\alpha>0$, so that
a probe is localized in the glass if it couples strongly enough.
As noted above, we will not discuss the weak-coupling limit where
the probe remains delocalized in the glass even for $\Fex=0$.

For non-zero $\Fex$, a further
bifurcation point will describe probe delocalization
in the glass through external force. It can be found by demanding that
the implicit-function theorem is violated for Eq.~\eqref{longtimetracer},
$\det\A^s=0$, leading to a
biquadratic equation,
\begin{multline}\label{fexbiq}
0 = \kappa_\perp^2\kappa_\parallel^2 (\Fexc)^4
  - \left(v_1^s f - 1\right)\left(\kappa_\perp^2 \left(v_1^s f + 1\right) +
    \kappa_\parallel^2\right) (\Fexc)^2 \\
  + \left(v_1^s f + 1\right) \left(\left(v_1^s f - 1\right)^2
  - (v_2^s f)^2\right)\,.
\end{multline}

\begin{figure}
\centerline{\includegraphics[width=\columnwidth]{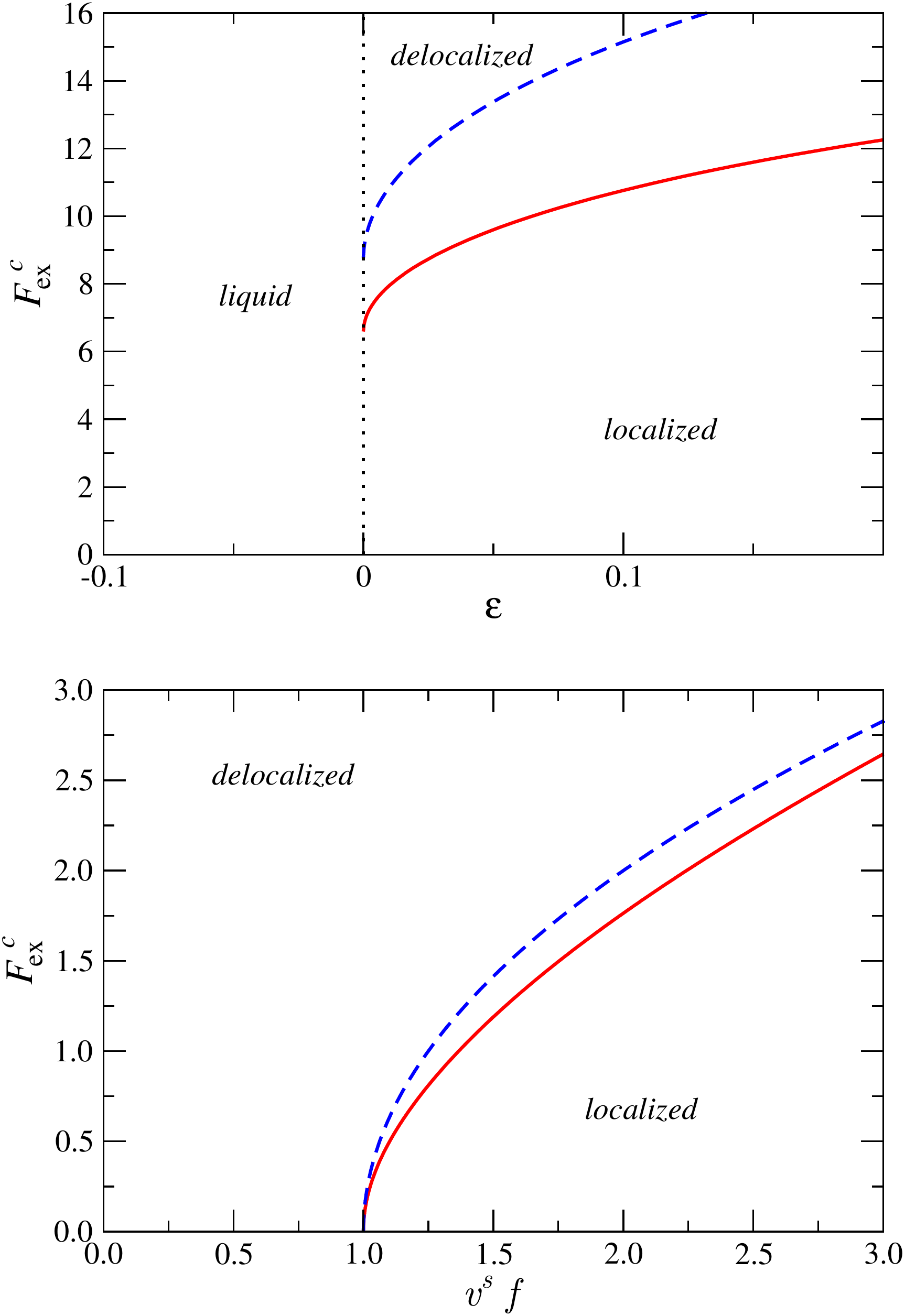}}
\caption{\label{fig:phasediag}
  State diagram for the probe particle in the schematic model.
  Upper panel: critical force $\Fexc$ as a function of distance to the
  host-liquid glass transition $\varepsilon$, for fixed $v^s$ as in
  Fig.~\ref{fig:corr_overview}. Lower panel: critical force as a function
  of probe-to-host coupling strength $v^sf$. Solid lines mark the boundary
  between localized and delocalized probe states; a vertical dotted line
  in the upper panel indicates the host-liquid glass transition.
  Dashed lines are $\Fexc$ obtained for the model without taking into account
  the anisotropy of the probe-particle fluctuations, $v_2^s=0$.
}
\end{figure}

Solving Eq.~\eqref{fexbiq} to obtain $\Fexc$ as a function of $v^sf$, we
obtain a decomposition of the parameter space for our schematic model.
The result for
exemplary parameters is shown in Fig.~\ref{fig:phasediag}. The lower panel
shows $\Fexc$ as a function of $v^sf$ directly. Typically, $v^sf$ will
change implicitly as the glass form factor $f$ changes with changing
$\varepsilon$; this variation is shown in the upper panel of the figure
for the parameters chosen as above.
In Fig.~\ref{fig:phasediag},
also the result after setting $v_2^s=0$ (as implicit in the
model of Ref.~\cite{Gazuz.2009}) is shown for
comparison (dashed lines).
For this aspect of the discussion, both models are very similar.
For $v^sf<1$, no real solution of Eq.~\eqref{fexbiq} exists, and the probe
particle is always delocalized. This is trivially the case in the liquid,
where $f=0$. In the glass, $f>0$ holds, and by assumption we restrict ourselves
to the case where the probe particle also becomes localized without any
external force, so that the jump from
zero to $f_c$ is big enough to render $v^sf_c>1$ at the glass transition.
As exemplified in the upper panel of Fig.~\ref{fig:phasediag}, this
leads to a nonvanishing critical force $\Fexc>0$ for the delocalization
of the probe at the glass transition, identifying the regime $\Fex<\Fexc$
as the one where the probe remains localized inside the glass even under
the action of an external force.
Beyond that, $f$ increases with $\varepsilon$, asymptotically as
$f-f_c\sim\sqrt{\varepsilon}$. Thus also $\Fexc$ increases in the glass.
A full analysis of Eq.~\eqref{fexbiq} shows that a further line of
solutions is present inside this localized regime. Based on the
following discussion of $\vec f^s$, we do not assign physical significance
to this.

\begin{figure}
\centerline{\includegraphics[width=\columnwidth]{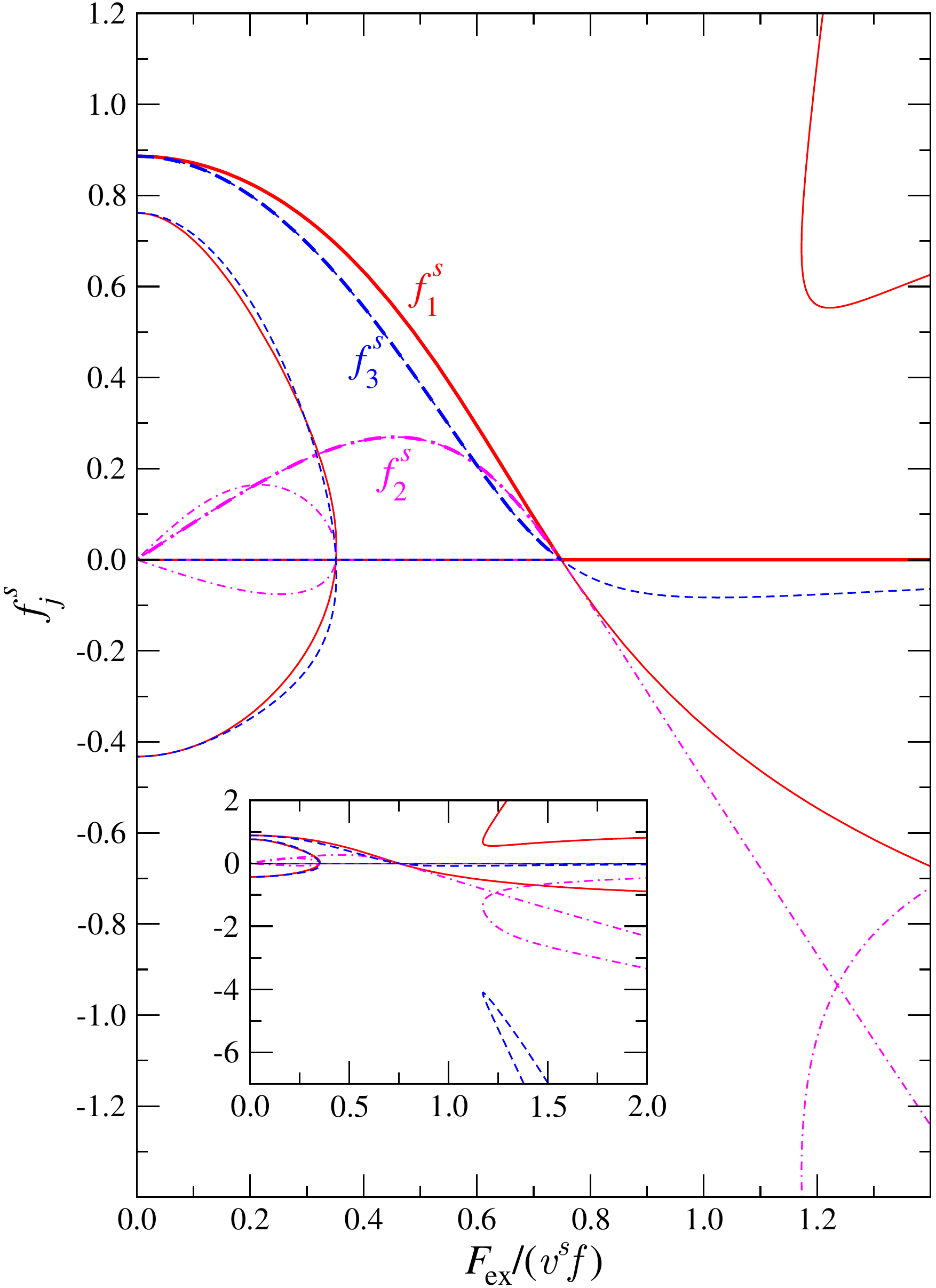}}
\caption{\label{fig:phasediagf} Solutions of the implicit set of
equations Eq.~\eqref{longtimetracer},
$f_1^s=\Real f^s_\parallel$ (solid lines),
$f_2^s=\Imag f^s_\parallel$ (dash-dotted), and $f_3^s=f^s_\perp$ (dashed),
as a function of the reduced applied force, $\Fex/(v^s f)$.
Parameters are $v^s=30$, $f=1-1/\sqrt2$ corresponding to the glass-transition
point of the $\text{F}_\text{12}$ model at $v_2^c=2$.
The solutions identified as the long-time limits
of the schematic model, Eq.~\eqref{sjoe}, are shown as thick lines.
The inset shows the full set of solution branches, including one for
$f_3^s$ that is cut off in the full figure for clarity.
}
\end{figure}

Solving Eq.~\eqref{longtimetracer} numerically, we obtain a set of
solutions to this implicit equation. These are shown in
Fig.~\ref{fig:phasediagf}, where we have highlighted those in bold that
we will use in the following as the physical ones.
For $\Fex<\Fexc$
they are the continuation of the known solution $f^s_\alpha(\Fex=0)$.
In particular, $f^s_2\to0$ as $\Fex\to0$ since the imaginary part of the
correlation function has to vanish in equilibrium, and $f^s_1=f^s_3$ for
$\Fex=0$.
The physical branch crosses zero at $\Fexc$, and for larger $\Fex$, no solution
emerges for which both real parts, $f^s_1$ and $f^s_3$, are non-negative
real, except $f^s_\alpha\equiv0$. We hence are dealing with a continuous
probe-delocalization transition: the probe-correlator nonergodicity parameters
$\vec f^s$ do not exhibit a jump at $\Fexc$.
Note that the choice of solution
branches is nontrivial: in standard equilibrium MCT, it can be proven
that the solution that is the largest nonnegative for \emph{all} components
has to be chosen \cite{Goetze.1995}. The proof relies on
mathematical properties of Eq.~\eqref{longtimebath} and its proper
generalizations, that are not ensured in general in Eq.~\eqref{longtimetracer}.
The number of solution branches to Eq.~\eqref{longtimetracer}
may depend on the parameters, and in particular the choice of the
$v_1^s$ and $v_2^s$. Fig.~\ref{fig:phasediagf} shows a case that is typical
for the description of microrheology data.

It can be seen that at $\Fex=\Fexc$, the largest real eigenvalue of
the matrix $\tensor A^s$ crosses zero, and that this eigenvalue is
always non-degenerate for the restriction to the parameters introduced
above. Thus the critical force is determined by a bifurcation in
Eq.~\eqref{longtimetracer} with a single critical direction
(co-dimension $d_c=1$). It is in this sense similar to the known
continuous delocalization transitions discussed within tagged-particle
models of MCT \cite{Goetze.2009}. These are quite different
from the standard MCT bifurcation, as we are not dealing with the
coalescence of two solution branches, but with the crossing of one
particular solution with the $f^s_j=0$ branch. $\Fexc$ is well-defined
since setting any of the $f^s_j=0$ in Eq.~\eqref{longtimetracer} also
demands that the other $f^s_{i\neq j}=0$.

For the later discussion we define reduced distances to the critical
points of the model, $\varepsilon$ and $\delta$. We set
\begin{subequations}
\begin{gather}
\vec{V} = (v_1,v_2)^{\top} = \vec{V}_{{c}} + \varepsilon \cdot \vec{b}_{{c}}\,,
\\
\vec{W} = (v_1^s,v_2^s,\Fex)^{\top} = \vec{W}^{{c}} + \delta \cdot \vec{b}^{s,{{c}}}\,.
\end{gather}
The vectors $\vec b_{c}$ and $\vec b^{s,{c}}$ can in prinicple be
chosen arbitrarily. In particular, the weak-coupling limit of small
$v^s$ is contained in the following derivation. For the sake of simplicity,
we will choose for all explicit calculations, in agreement
with above,
\begin{align} \vec b_{c}&=(v_{1,c},v_{2,c})^\top\,, &
  \vec b^{s,{c}}&=(0,0,\Fexc)^\top\,. \end{align}
\end{subequations}
With this choice, $\delta$ becomes a reduced force, such that
$\delta<0$ corresponds to $\Fex<\Fexc$, i.e., the localized regime,
and $\delta>0$ to the delocalized regime.
Recall that subscripts $c$ refer to quantities evaluated at $\varepsilon=0$;
we reserve superscripts $c$ for quantities evaluated at $\delta=0$.

\section{The beta-scaling law}\label{beta}

\subsection{Scaling Functions}

Having identified the long-time limits of the correlation functions in the
glassy-host, localized-probe regime, we now turn to a discussion of the
dynamical correlation functions $\phi(t)$ and $\phi^s_\alpha(t)$. We
anticipate that upon approaching a critical point, $\varepsilon\to0$
for small $\delta$, the correlators
will stay close to an intermediate plateau given by $f$
over a time window that increases
as the distance parameters approach zero.
Thus, $\phi(t)-\tilde f$ and $\vec\phi^s(t)-\tilde{\vec f}^s$ can be
identified as small parameters on a dynamical time scale
that will be determined below. Here, $\tilde f$ is a parameter introduced
in Ref.~\cite{Goetze1991a}, approximating
$f$ for $\varepsilon\ge0$ and continuing the latter smoothly to $\varepsilon<0$
such that $\lim_{\varepsilon\to\pm0}\tilde f=f_c$. For the
$\text{F}_\text{12}$ model, where we only consider $\mathcal A_2$ bifurcations,
$\tilde f$ is given by the unique positive solution of
$v_1+2v_2\tilde f-1/(1-\tilde f)^2=0$ that obeys $\tilde f=f_c$ at the
critical point. We define $\tilde{\vec f}^s$ and
a corresponding matrix $\tilde{\tensor A}^s$ by
Eqs.~\eqref{longtimetracer} and \eqref{longtimetracerA}
with $f$ replaced by $\tilde f$.

For the host correlator $\phi$ given by Eqs.~\eqref{f12},
the asymptotic analysis has been worked
out in detail. Splitting $\phi(t)=\tilde f+G(t)$,
there holds an expansion in terms of the small parameter
$\sigma$: $G(t)=c_\sigma hg_\sigma(t)+\mathcal O(c_\sigma^2)$ with
$c_\sigma=\sqrt{|\sigma|}=\mathcal O(\sqrt{|\varepsilon|})$. The
leading order correction to the plateau is called $\beta$-correlator
$g_\sigma(t)$.
There appears a critical amplitude $h$, connected to the critical eigenvector
arising in the bifurcation analysis of Eq.~\eqref{longtimebath} -- in the
one-correlator schematic model, this is just a prefactor, set to
$h=(1-f_c)$ by convention.
The function $g(t)$
describes relaxation to and from the plateau (visible e.g.\ in
Fig.~\ref{fig:corr_overview}) as asymptotic power laws,
\begin{subequations}
\begin{align}
  g_\sigma(t)\propto(t/t_0)^{-a} & &\text{for $t_0\ll t\ll t_\sigma$,}\\
  g_\sigma(t)\propto-(t/t_\sigma)^b & &\text{for $t_\sigma\ll t\ll t_\sigma'$,}
\end{align}
\end{subequations}
where the latter, called the von~Schweidler law, only occurs inside the
liquid, $\varepsilon<0$. Here, diverging time scales have been identified,
\begin{align}\label{bathtimescales}
  t_\sigma&=t_0/|\sigma|^{1/(2a)}\,, &
  t_\sigma'&=t_0/|\sigma|^{1/(2a)+1/(2b)}\,,
\end{align}
where $\sigma$ is a distance parameter obeying $\sigma\sim\varepsilon$
as $\varepsilon\to0$. For the schematic model and our definition of
$\varepsilon$, one gets $\sigma=\varepsilon f^c(1+f^c)(1-f^c)
\approx0.38\varepsilon$.
The microscopic time scale $t_0$
is fixed by the short-time motion of the correlation function. The
power-law exponents $a,b>0$ are solutions of
\begin{equation}
  \frac{\Gamma(1-a)^2}{\Gamma(1-2a)} = \frac{\Gamma(1+b)^2}{\Gamma(1+2b)}
  =\lambda\,,
\end{equation}
with the exponent parameter $\lambda$ given by $\lambda = 1/\sqrt{v_2}$ on the
bifurcation manifold \cite{Goetze.2009}.

Introducing a rescaled time, $\hat t=t/t_\sigma$, the $\beta$ correlator
is found to obey the $\beta$-scaling equation written with the scaled Laplace
frequency $\hat z=zt_\sigma$ as
\begin{equation}\label{betascaling}
  \mp1+\lambda\hat z\widehat{g_\pm^2}(\hat z)+(\hat z\hat g_\pm(\hat z))^2=0\,,
\end{equation}
where $g_\pm(\hat t)$ are the scaling solutions for $\sigma\gtrless0$
that depend only on the sign of $\sigma$.
They obey $g_\pm(\hat t)\sim{\hat t}^{-a}$ for $\hat t\to0$, and
$g_-(\hat t)\sim-B\hat t^b$ for $\hat t\to\infty$. The asymptotic form
of the relaxation to and from the plateau thus is given by a scaling law
with only $\lambda$ as a non-universal parameter.
The positive constant $B=\mathcal O(1)$ depends on $\lambda$ and has
been tabulated \cite{Goetze.1990}.
For $\varepsilon>0$, $g_+(\hat t)$ approaches a constant as $\hat t\to\infty$,
giving the correction to the critical plateau value.

In the same spirit we now write
\begin{equation}
  \vec\phi^s(t)=\tilde{\vec f}^s+c_\sigma\vec g_\sigma^s(\hat t)
  +\mathcal O(c_\sigma^2)
  \,.
\end{equation}
To arrive at an equation for $\vec g_\sigma^s$ that is asymptotically
valid for $\sigma\sim\varepsilon\to0$, we follow the standard procedure
laid out in Ref.~\cite{Goetze.2009}: introducing
$\hat z=zt_\sigma$, we collect in Eq.~\eqref{sjoephilt} the
leading terms in $c_\sigma$, making use of Eq.~\eqref{longtimetracer}.
We arrive at
\begin{align}\label{expansioneps}
  0 =& \left\{\tilde\A^s\cdot\hat z\hat{\vec g}^s_\sigma(\hat z)
  - \tilde f\vec{\mathcal D}^s[\tilde{\vec f}^s,\hat z\hat{\vec g}_\sigma^s(\hat z)]
  \right.\nonumber\\ &\left.
  + \hat z\hat g_\sigma(\hat z)
    \vec{\mathcal C}^s[\tilde{\vec f}^s,\vec\phi_0^s-\tilde{\vec f}^s]\right\}
  \nonumber\\ &+ c_\sigma\left\{\vec{\mathcal C}^s[
    \hat z\widehat{g_\sigma\vec g_\sigma^s}(\hat z),\vec\phi_0^s-\tilde{\vec f}^s]
  +\hat z\hat g_\sigma(\hat z)
    \vec{\mathcal C}^s[\tilde{\vec f}^s,\hat z\hat{\vec g}^s_\sigma(\hat z)]
  \right.\nonumber\\ &\left.\phantom{c_\sigma\Big\{}
  +\tilde f\vec{\mathcal C}^s[\hat z\hat{\vec g}^s_\sigma(\hat z),
    \hat z\hat{\vec g}^s_\sigma(\hat z)]\right\}
  +\mathcal O(c_\sigma^2,(c_\sigma t_\sigma)^{-1})
\end{align}
Note that since the exponent $a<1/2$ \cite{Goetze.2009}, the dropped terms
$(c_\sigma t_\sigma)^{-1}\sim|\sigma|^{1/(2a)-1/2}$ are indeed of higher order.

The leading order for $\varepsilon\to0$ in Eq.~\eqref{expansioneps} results
in a linear equation system,
\begin{equation}\label{betadeltafar0}
\tensor{L}^s_c \cdot \vec{g}_{\pm}^s\left(\hat{t}\right) = g_{\pm}\left(\hat{t}\right) \vec{l}^s_c\,,
\end{equation}
where the linear mapping $\tensor L^s_c$ is defined through
$
\tensor L^s_c=\A^s_{c}-f_c\vec{\mathcal D}^s_c
[\vec f^s_c,\cdot]
$
and the inhomogeneity reads
$
\vec{l}^s_c = -\vec{\mathcal C}^s[\vec{f}^s_c,\vec\phi_0^s-\vec{f}^s_c]\,.
$
For state points far enough from the critical force, Eq.~\eqref{betadeltafar0}
determines $\vec g^s_\pm(\hat t)$ by the host liquid $g_\pm(\hat t)$. This
expresses that in such cases, the asymptotic dynamics of the tracer is
governed by that of the host liquid, i.e.,
\begin{subequations}\label{betacoupling}
\begin{equation}\label{betacouplingg}
  \vec g_\pm^s(\hat t)=\vec h^s_cg_\pm(\hat t)\,,
  \qquad\text{$\delta\neq0,$}
\end{equation}
with an amplitude given by
\begin{equation}
  \vec h^s_c=-{\tensor L^s_c}^{-1}\vec{\mathcal C}^s[\vec f^s_c,
  \vec\phi_0^s-\vec f^s_c]\,.
\end{equation}
\end{subequations}

\begin{figure}
\centerline{\includegraphics[width = \columnwidth]{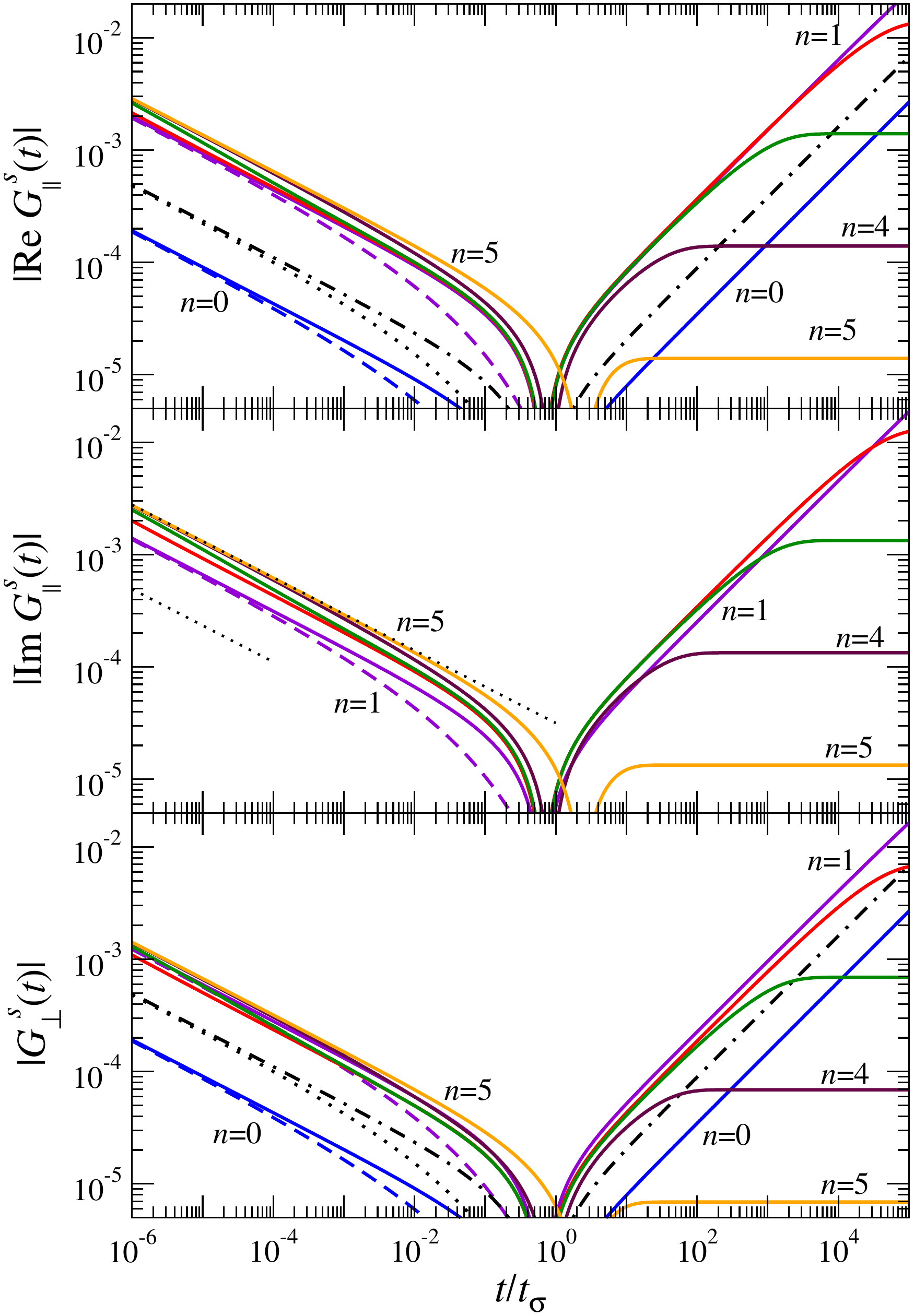}}
\caption{\label{fig:corr_beta_2}
  Correlation functions $\vec\phi^s(t)$
  close to the glass transition point, for various
  external forces, plotted as the leading-order deviation from the plateau
  value,
  $|\vec G^s(t)|=|\vec\phi^s(t)-\vec f^s|$ for $\varepsilon\to0$, as
  a function of $t/t_\sigma$.
  Parameters as in Fig.~\ref{fig:zeta_overview};
  solid lines correspond to liquid states, $\varepsilon=-2.5\times10^{-10}$,
  dashed lines to glassy states, $\varepsilon=2.5\times10^{-10}$. Various
  forces $\Fex$ are shown with $\delta=(\Fex-\Fexc)/\Fexc=10^{-n}$,
  $n=0$, $1$, $2$, $3$, $4$, and $5$ as labeled. (For the glass,
  only $n=0$ and $n=1$ are shown.)
  Dash-dotted and dotted lines indicate the corresponding host $\beta$-correlator
  $|G(t)|$, for the liquid respectively the glass, in the panels showing
  the real parts. For the imaginary part, dotted lines indicate
  the critical law of the host correlator, $G(t)\sim h(t/t_0)^{-a}$,
  and $h^s_{c,j}/\lambda\approx3.985/\lambda$ times this critical law.
}
\end{figure}

In Fig.~\ref{fig:corr_beta_2}, we show exemplary results for the correlators
$\vec\phi^s(t)$ close to the glass transition, in a double-logarithmic plot
of $|\vec\phi^s(t)-\vec f^s|$
to exhibit the asymptotic power laws. Two states corresponding to
small $|\varepsilon|$ were chosen that exhibit a large window of validity
for the asymptotic Eq.~\eqref{expansioneps}. As dash-dotted and dotted
lines, the corresponding $\phi(t)$ are shown for $\varepsilon<0$ and
$\varepsilon>0$, respectively. They follow the asymptotic critical power law,
$\phi(t)-f^c\sim h(t/t_0)^{-a}$ over several decades in time; a numerical
estimate yields $t_0\approx0.146$ for the parameters chosen here.
The $\varepsilon<0$ curve
also exhibits von~Schweidler's law, $\phi(\hat t)-f^c\sim-B{\hat t}^b$
at large rescaled times $\hat t=t/t_\sigma$. With the distance parameter
chosen here, we get $\sigma=6.7\times10^{-11}$ and consequently
$t_\sigma=8.07\times10^{14}$; the parameter $B$ is tabulated
\cite{Goetze.1990} and estimated as $B=0.68$.

For forces far from the critical threshold, e.g., $\delta=-1$,
Eq.~\eqref{betacoupling} is valid. This is demonstrated by the $n=0$ curves
in Fig.~\ref{fig:corr_beta_2} that correspond to the force-free case:
the probe-particle correlation functions
closely following the host-liquid curves over all time, up to a fixed
amplitude. Evaluation of Eq.~\eqref{betacoupling} in this case yields
$\vec h^s_c\approx(0.39,0,0.39)^\top$, which is easily verified in the figure.
In this sense, the tagged-particle dynamics is governed by
the dynamics of the host liquid. This well known fact is the basis
for various approximations relating single-particle motion
and collective dynamics close to the glass transition.
For $n=1$, we get $\vec h^s_c\approx(3.99,2.83,2.54)^\top$, and this is
also verified by Fig.~\ref{fig:corr_beta_2}.

Approaching $\Fexc$, shown by curves with $\delta=-10^{-n}$ closer to
zero (larger $n$) in Fig.~\ref{fig:corr_beta_2}, the coupling found for
$\delta=0$ breaks down. Most obviously, in the regime of the von~Schweidler
law, $\hat t\gg1$, the curves for $n\neq0$ decay to zero more rapidly
than the host correlator, indicated by a plateau in Fig.~\ref{fig:corr_beta_2}
(dashed lines). This regime will be discussed later.

More subtly, for $\hat t\ll1$, the probe correlators for $\Fex\neq0$ still
appear proportional to the host liquid correlator, but with a different
prefactor than the one given in Eq.~\eqref{betacoupling}. For
$n=5$, Eq.~\eqref{betacoupling} yields $\vec h^s_c\approx(4.16,3.98,2.05)$,
not compatible with the asymptotes seen in Fig.~\ref{fig:corr_beta_2}.
We address
this now, discussing the limit $\delta\to0$ of Eq.~\eqref{expansioneps}.

At the critical force, $\delta=0$, we have $\det\A^{s,c}=0$ and concomitantly
$\lim_{\delta\to0}\vec f^s_c=\vec f^{s,c}_c=0$, as discussed above. In that
case, also $\det\tensor L^{s,c}_c=0$, rendering Eq.~\eqref{betadeltafar0}
singular and Eq.~\eqref{betacoupling} invalid.
In leading order in $\delta$, $\vec g^s_\sigma(t)$ must be in the kernel
of $\det\A^{s,c}$. In our model, this matrix is only simply degenerate,
and we denote by $\vec h^{s,c}$ a corresponding null-eigenvector.
Keeping the ansatz that $\vec g^s_\sigma(t)$ depends linearly on
$g_\sigma(t)$, we modify Eq.~\eqref{betacouplingg},
\begin{equation}\label{ansatz2}
  \vec g^s_\sigma(\hat t)=\vec h^{s,c}g_\sigma(\hat t)
  +\mathcal O(c_\sigma,\delta)\,,\qquad\text{$\delta\to0$.}
\end{equation}
Differentiating Eq.~\eqref{longtimetracer} with respect to $\delta$
at $\delta=0$, again using that $\tilde{\vec f}^{s,c}=0$, we find
$\tilde\A^{s,c}(d/d\delta)\tilde{\vec f}^{s,c}=0$ and hence
$(d/d\delta)\tilde{\vec f}^{s,c}=\eta\vec h^{s,c}$ with some
constant $\eta$.
Here and in the following, expressions like $d/d\delta\,\vec f^{s,c}$ are
to be read as $\lim_{\delta\to0}d/d\delta\,\vec f^s$.

Accounting for
the next-to-leading order terms in Eq.~\eqref{expansioneps} and using
$\tilde f_j^{s,c}=0$, we get from taking the derivative with respect
to $\delta$
\begin{multline}\label{expansionepsdel}
  0=(\hat z\hat g_\sigma(\hat z))\,\delta\,
  \times\\ \times
  \left\{\frac{d\tilde\A^{s,c}}{d\delta}\vec h^{s,c}
  -\eta\tilde f\vec{\mathcal D}^{s,c}[\vec h^{s,c},\vec h^{s,c}]
  +\eta\vec{\mathcal C}^{s,c}[\vec h^{s,c},\vec\phi_0^s]\right\}\\
  +c_\sigma\left\{\vec{\mathcal C}^{s,c}[\vec h^{s,c},\vec\phi_0^s]
  \hat z\widehat{g_\sigma^2}(\hat z)
  +\tilde f\vec{\mathcal C}^{s,c}[\vec h^{s,c},\vec h^{s,c}]
  (\hat z\hat g_\sigma(\hat z))^2\right\}
  \\
  +\mathcal O\left(c_\sigma^2,(c_\sigma t_\sigma)^{-1},
    \delta^2,c_\sigma\delta\right)
\end{multline}
In the following, we will suppress the indication of omitted higher-order
terms as far as they are already denoted here.
The term containing $(d/d\delta)\tilde\A^{s,c}$ can be eliminated in
favor of $\vec{\mathcal C}^{s,c}$ by looking at the Taylor expansion
of Eq.~\eqref{longtimetracer}, which yields
\begin{multline}
  0=\tilde\A^{s,c}\cdot\frac12\frac{d^2\tilde{\vec f}^{s,c}}{d\delta^2}
  +\eta\frac{d\tilde\A^{s,c}}{d\delta}\cdot{\vec h}^{s,c}\\
  -\tilde f\eta^2\vec{\mathcal C}^{s,c}[{\vec h}^{s,c},{\vec h}^{s,c}]\,.
\end{multline}
Multiplying with the left null-eigenvector $\hat{\vec h}^{s,c}$ of
$\tilde\A^{s,c}$, the first term vanishes, and inserting into
Eq.~\eqref{expansionepsdel} provides
\begin{multline}\label{expandetermineA}
  0=(\hat z\hat g(\hat z))\,\eta\delta\,
  \left\{\hat{\vec h}^{s,c}\vec{\mathcal C}^{s,c}[\vec h^{s,c},\vec\phi^s_0]
  -\tilde f\hat{\vec h}^{s,c}\vec{\mathcal C}^{s,c}[\vec h^{s,c},\vec h^{s,c}]
  \right\}\\
  +c_\sigma\left\{\hat{\vec h}^{s,c}\vec{\mathcal C}^{s,c}[\vec h^{s,c},
    \vec\phi^s_0]\hat z\widehat{g_\sigma^2}(\hat z)
  \right.\\ \left.
    +\tilde f\hat{\vec h}^{s,c}\vec{\mathcal C}^{s,c}[\vec h^{s,c},
    \vec h^{s,c}](\hat z\hat g_\sigma(\hat z))^2\right\}
\end{multline}

\begin{figure}
\centerline{\includegraphics[width = \columnwidth]{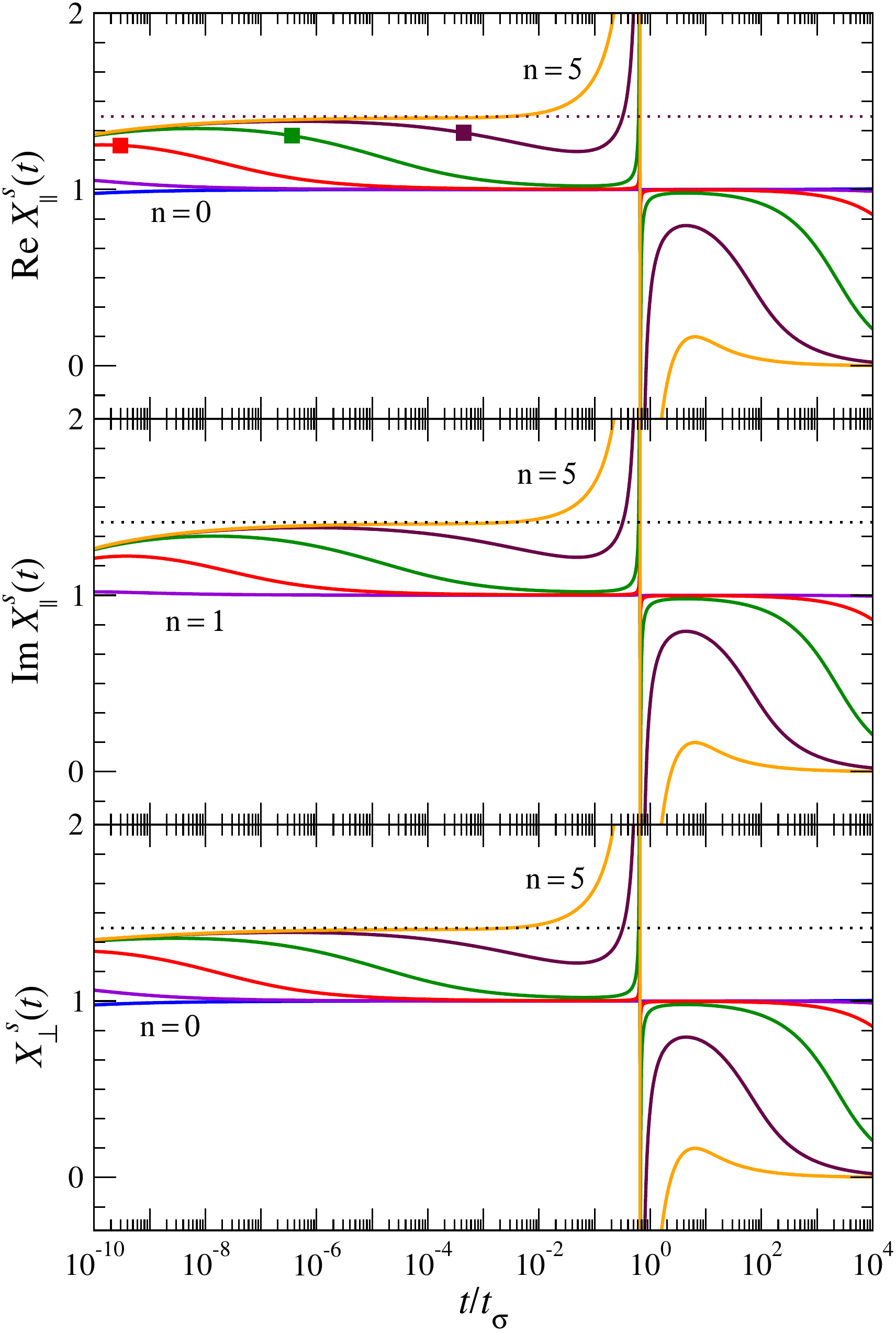}}
\caption{\label{fig:corr_beta_1}
  Normalized ratio of probe-correlation functions to that of the host liquid,
  $X_j(t)=[\phi_j^s(t)-\tilde f_j^s]/[\phi(t)-\tilde f]/(u_j^{s,c}A_\delta)$,
  cf.\ Eq.~\eqref{adelta}, as a function of $\hat t=t/t_\sigma$.
  Parameters are chosen as in Fig.~\ref{fig:corr_beta_2}; only liquid
  curves, $\varepsilon=-2.5\times10^{-10}$ are shown, for
  $\delta=-10^{-n}$ with $n=0$, $1$, $2$, $3$, $4$, and $5$ (indicated
  by the labels).
  The dotted line indicates $A_\varepsilon/A_\delta=\lambda=\sqrt2$.
  Squares in the upper panel indicate the time scale $t_\delta$.
}
\end{figure}

Let us discuss the relevant limits of Eq.~\eqref{expandetermineA}.
We assume that $\vec{\mathcal C}^{s,c}[\vec h^{s,c},\vec h^{s,c}]\neq0$,
which is generically the case.
Note that in writing Eq.~\eqref{ansatz2}, the normalization of the
eigenvector $\vec h^{s,c}$ is left undetermined. In the limit
$\varepsilon\to0$, this normalization is now fixed by requiring the
first bracket in Eq.~\eqref{expandetermineA} to vanish. If we introduce through
$\vec h^{s,c}=A\vec u$ the normalized eigenvector $\vec u$ and the
amplitude $A=\|\vec h^{s,c}\|$, we obtain
\begin{equation}\label{adelta}
  A_\delta
  =\frac{\hat{\vec h}^{s,c}\cdot\vec{\mathcal C}_c^{s,c}[\vec u,\vec\phi^s_0]}
  {f_c\hat{\vec h}^{s,c}\cdot\vec{\mathcal C}^{s,c}[\vec u,\vec u]}
\end{equation}
where the subscript recalls that this result is valid for taking the
limit $\delta\to0$ after $\varepsilon\to0$. The amplitude $A_\delta$
does not depend on the actual path along which the hypersurface of
delocalization transitions is crossed.

Taking $\delta\to0$ first, we require the second bracket in
Eq.~\eqref{expandetermineA} to vanish. Letting $\varepsilon\to0$ after
that, we can further make use of Eq.~\eqref{betascaling} to
arrive at
\begin{equation}\label{aepsilon}
  A_\varepsilon
  =\frac{\hat{\vec h}^{s,c}\cdot\vec{\mathcal C}_c^{s,c}[\vec u,\vec\phi^s_0]}
  {\lambda f_c\hat{\vec h}^{s,c}\cdot\vec{\mathcal C}^{s,c}[\vec u,\vec u]}
  =\frac{A_\delta}\lambda\,.
\end{equation}

Equations \eqref{adelta} and \eqref{aepsilon} together with
Eq.~\eqref{ansatz2} imply that one recovers both the critical power law,
$\vec g^s_\pm(\hat t)\sim{\hat t}^{-a}$, and the von~Schweidler law,
$\vec g^s_-(\hat t)\sim-{\hat t}^b$, in the double limit $\varepsilon\to0$
and $\delta\to0$. The ratio of the correlation functions, however, depends
on the order in which this double limit is taken. To exemplify this,
define the ratio
$X^s_j(t)=(\phi^s_j(t)-\tilde f^s_j)/([\phi(t)-\tilde f]A_\delta u_j)$.
Results for typical parameters are shown in Fig.~\ref{fig:corr_beta_1}.
According to Eqs.~\eqref{adelta} and \eqref{aepsilon}, the ratio approaches
unity for $\hat t\to\infty$ at not too small $\delta$.  This is demonstrated
by the $n=0$ curve in the figure. Lowering $\delta\to0$, one notices a second
plateau in $X^s_j(t)=1/\lambda$ that becomes more pronounced for smaller
$|\delta|$. For Fig.~\ref{fig:corr_beta_1},
the result is only valid for $t<t_\sigma$. The ranges of validity
of the various scaling predictions will be discussed below.

Another scaling result emerges if instead of Eq.~\eqref{ansatz2} we consider
the case that $g_\sigma(\hat t)$ is of higher order and can be set to zero
in Eq.~\eqref{expansioneps}. We will identify a time scale where this is
admissible below. Requiring the leading order in Eq.~\eqref{expansioneps}
to vanish again leads to the requirement $\hat{\vec g}_\sigma^s
\propto\vec h^{s,c}$, the null-eigenvector of $\tilde\A^{s,c}$.
The expansion Eq.~\eqref{expansionepsdel} reduces to
\begin{multline}\label{equmotbathzero}
  0 =\delta\left\{\frac{d\tilde\A^{s,c}}{d\delta}\cdot
    \hat z\hat{\vec g}^s_\sigma(\hat z)
  -\tilde f{\mathcal D}^{s,c}\left[
    \frac{d\tilde{\vec f}^{s,c}}{d\delta},\hat z\hat{\vec g}^s_\sigma(\hat z)\right]
  \right\}
  \\
  +c_\sigma
  \tilde f\vec{\mathcal C}^{s,c}[\hat z\hat{\vec g}^s_\sigma(\hat z),
    \hat z\hat{\vec g}^s_\sigma(\hat z)]
   +\mathcal O\left(c_\sigma^2,(c_\sigma t_\sigma)^{-1},
    \delta^2,c_\sigma\delta\right)
\end{multline}
The limit $\varepsilon\to0$ is not meaningful in this equation, since
the first bracket then only admits the trivial solution. Letting, however,
$\delta\to0$, 
we gain $\vec{\mathcal C}^{s,c}[\hat z\hat{\vec g}
_\sigma^s(\hat z),\hat z\hat{\vec g}_\sigma^s(\hat z)]=0$, which is
consistent if we set
\begin{equation}\label{decay12}
 \vec{g}_{\sigma}^s(\hat{t}) = C \hat{t}^{-1/2} {\vec{h}}^{s,c}\,.
\end{equation}
The constant $C$ has to be determined by matching the various
asymptotic expansions.

\subsection{Time Scales}

The different results derived above present the asymptotic behavior of the
probe-particle correlation functions in various limits. These correspond
to different time scales for which the results hold, which we discuss now.
Recall the Laplace transform of a power law
$g(t)=t^{-x}$ yields $z\hat g(z)=-\Gamma(1-x)(-iz)^x$.
For short rescaled times, $\hat t\to0$, the host-liquid $\beta$ correlator
assumes the form $g_\sigma(\hat t)\sim{\hat t}^{-a}$. In
Eq.~\eqref{expandetermineA}, we then identify a time scale $\hat t_\delta$
that separates the two scaling limits discussed in connection with
Eqs.~\eqref{adelta} and \eqref{aepsilon}: for $\hat t\ll\hat t_\delta$,
the second curly bracket in Eq.~\eqref{expandetermineA} dominates,
and we recover the $\delta\to0$ regime. For $\hat t\gg\hat t_\delta$,
the first curly bracket dominates, corresponding to the $\varepsilon\to0$
regime. Balancing the power-law exponents, we get
$\hat t_\delta=\mathcal O(1)|\sigma/\delta^2|^{1/(2a)}$, or
\begin{equation}\label{tdelta}
  t_\delta=\mathcal O(t_0)|\delta|^{-1/a}\,.
\end{equation}

\begin{figure}
\centerline{\includegraphics[width = \columnwidth]{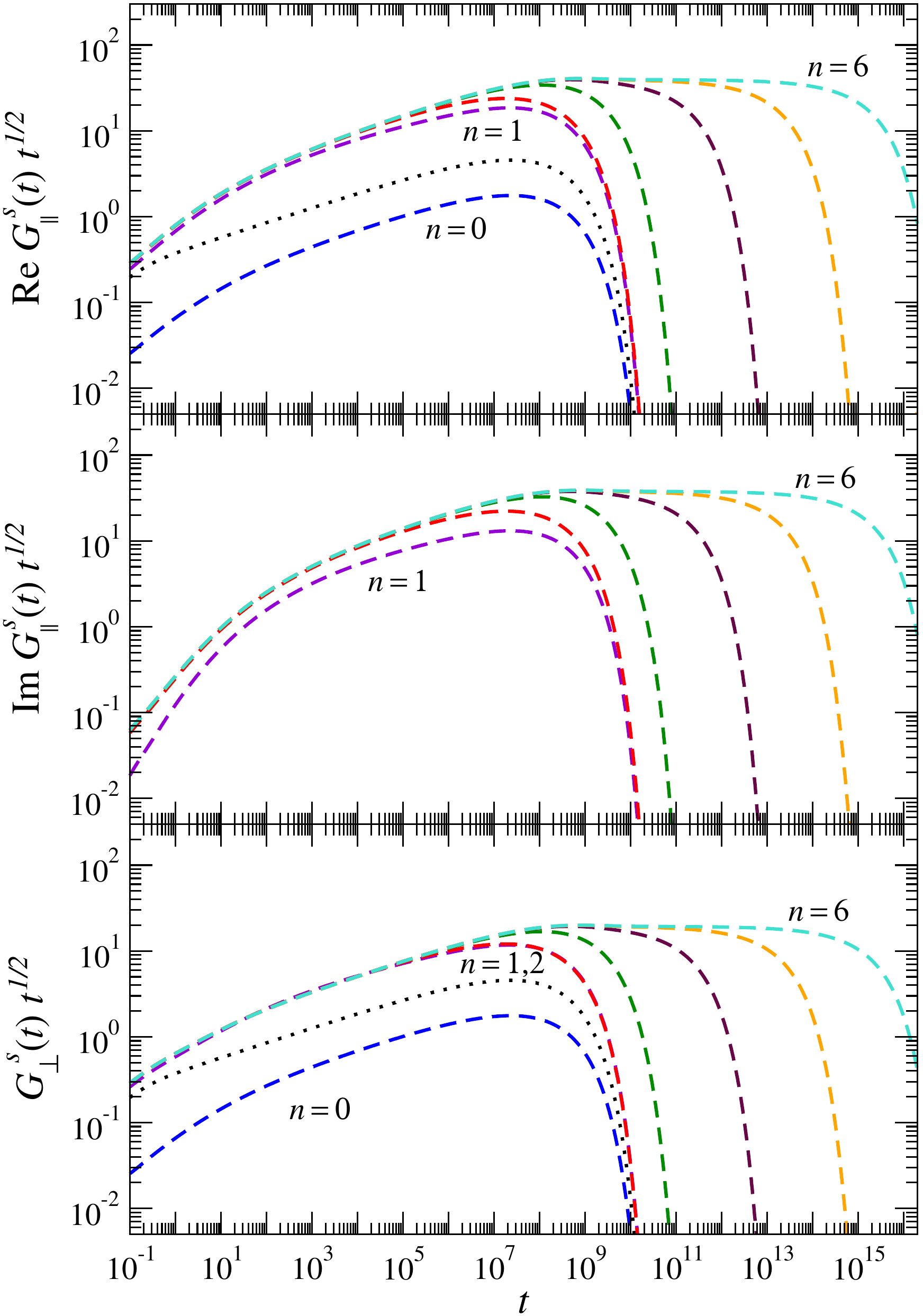}}
\caption{\label{fig:corr_beta_3} Correlation functions $\vec\phi^s(t)
-\vec f^s$
multiplied by $\sqrt t$, for $\varepsilon=10^{-6}$, and other parameters
as in Fig.~\ref{fig:corr_beta_2}. Reduced forces are
$\delta=-10^{-n}$ with $n=0,\ldots 6$. The dash-dotted line shows
the corresponding host-liquid correlator.
}
\end{figure}

Considering the case where $t_\delta\gg t_\sigma$, we have to distinguish
the two signs of $\varepsilon$. In the glass, $\varepsilon>0$, $g_\sigma$
attains a long time limit of $\mathcal O(\sqrt\sigma)$ so that we may
consider Eq.~\eqref{equmotbathzero} to be valid on a time scale
$\hat t_{1/2}=\mathcal O(1)|\sigma/\delta^2|$, or
\begin{equation}\label{t12}
  t_{1/2}=\mathcal O(t_0)|\sigma|^{1-1/(2a)}|\delta|^{-2}\,.
\end{equation}
For times $t\gg t_{1/2}$, the trivial solution of Eq.~\eqref{equmotbathzero}
is obtained. Hence the probe $\beta$ correlators $\vec G^s(t)$ show a
nonalgebraic decay to zero for times $t\sim t_{1/2}$. Matching this
solution, Eq.~\eqref{decay12}, to the asymptote obtained above, we find
that the constant $C$ obeys $C=C'\cdot|\sigma|^{1/2-1/(4a)}$, where
$C'$ is independent of $\sigma$ and $\delta$. This yields the scaling law
\begin{multline}\label{scalingcorr12}
  \vec G^s(t)\propto|\sigma|^{-1/(4a)}|\delta|(t/t_{1/2})^{-1/2}\,,
  \\ t_\sigma\ll t\ll t_{1/2}\,.
\end{multline}
We verify this scaling law in Fig.~\ref{fig:corr_beta_3}: for
$\varepsilon>0$ much larger than considered in Fig.~\ref{fig:corr_beta_2},
$\vec G^s(t)\sqrt t$ is seen to approach a constant at times long compared
to the ultimately exponential relaxation \cite{Goetze.1995} of $G(t)$.

In the liquid, $\varepsilon<0$, we can balance terms of $\mathcal O(\delta)$
with those of $\mathcal O(c_\sigma)$ inserting the von~Schweidler law
for the host-liquid $\beta$ correlator, $g_\sigma(\hat t)\sim -\mathcal O(1)
{\hat t}^{-b}$, provided that $\hat t\gg1$.
As a result, this asymptotic power law can be seen in the
probe correlator on a time scale
\begin{equation}\label{tsigdelta}
  t_{\sigma,\delta}'=\mathcal O(t_0)|\delta|^{1/b}
  |\sigma|^{-1/(2a)-1/(2b)}\,.
\end{equation}

Taken together, Eqs.~\eqref{bathtimescales}, \eqref{tdelta},
\eqref{t12}, and \eqref{tsigdelta} define five time scales determining
the asymptotic behavior of the $\beta$ correlation functions for the
probe pulled by an external force.
Two of them are inherited from the glass-transition dynamics of the host
liquid; $t_\sigma$ determines the time scale for the relaxation around
the plateau, while $t_\sigma'$ sets the time scale for the final decay
in the liquid.
These time scales are marked as diamonds in Fig.~\ref{fig:corr_overview_2};
we have seen that it suffices to study the correlator $\Real\phi^s_\parallel(t)$
in the asymptotic regime, since the other $\phi^s_j(t)$ are connected via
the critical amplitudes $\vec h^{s,c}$ to the same time-dependent laws.

The time scale $t_\delta$ (squares in Fig.~\ref{fig:corr_overview_2})
separates the regime that is dominated by
the proximity to the glass transition ($\sigma\to0$) from the one
dominated by the proximity to the delocalization transition ($\delta\to0$).
The curve for $\delta=-0.01$ exemplifies the case where $t_\delta\ll t_\sigma$
so that the dynamics for $t\gg t_\delta$ is dominated by the host-liquid glass
transition. In this case, the probe-particle $\beta$ correlator is proportional
to the host-liquid one, with an amplitude that changes by a factor
$1/\lambda$ at $t\approx t_\delta$. For $t\ll t_\delta$,
Eq.~\eqref{aepsilon} holds for this amplitude, while at $t\gg t_\delta$
it drops to the value determined by Eq.~\eqref{adelta}.
This explains the behavior of the correlator ratio $\vec X(t)$ shown
in Fig.~\ref{fig:corr_beta_1}. There, squares mark $t_\delta$; for the
particular cases shown, $t_\delta<t_\sigma$ holds, so that the ratio
for $t>t_\sigma$ is given by Eq.~\eqref{adelta}, viz.\ $X_j(t)=1$, as long
as the probe correlators do not decay to zero.

At times $t\gg t_\sigma\gg t_\delta$, the probe-particle correlator exhibits
force-induced decay that couples to the von~Schweidler law of the
host liquid: this is exemplified for $\delta=-0.01$ in
Fig.~\ref{fig:corr_overview_2} for $t\approx t_{\sigma,\delta}'$ (marked
by a triangle). There holds $t_{\sigma,\delta}'\propto|\delta|^{1/b}$, so
that approaching the delocalization threshold, $\delta\to0$, this
force-induced decay is accelerated according to a non-trivial power law.

The curve for $\delta=10^{-5}$ in Fig.~\ref{fig:corr_overview_2}
on the other hand exhibits
the case $t_\delta\gg t_\sigma$, and the dynamics for $t\ll t_\delta$ is
dominated by the probe-delocalization limit. In this case, the
time scale $t_{1/2}$ becomes relevant (marked by a circle in the figure),
and the probe correlator decays according to a power law.

Both $t_\delta$ and $t_{1/2}$ diverge
as the force threshold is approached, $\delta\to0$, while the host-liquid
time scales $t_\sigma$ and $t_\sigma'$, as well as the time scale for
which the probe motion couples to the host's von~Schweidler law,
$t_{\sigma,\delta}'$, diverge for $\varepsilon\to0$, i.e., upon approaching
the glass transition.
From the power laws, one infers the relevant scaling combination:
for $|\sigma/\delta^2|\gg1$, the dynamics is determined by the force-induced
delocalization, while for $|\sigma/\delta^2|\ll1$, the glass-transition
dynamics describes the relaxation.



\section{The $\boldsymbol\alpha$-scaling law}\label{alpha}

In the liquid, the decay of the host correlation function from the plateau
towards zero can be discussed on the time scale $t_\sigma'$ by another
scaling law, known as the $\alpha$-scaling law. There, a master function
can be derived for the limit $\sigma\to0$ in the form \cite{Goetze.2009}
\begin{equation}
  \phi(t)\asymp F(t/t_\sigma')\,.
\end{equation}
The function $F$ is independent of $\sigma$. As a result, the final relaxation
of the correlators displays a superposition principle, allowing the long-time
part of the correlators to be scaled to a master curve by scaling time
with $t_\sigma'$. We derive a similar law for the probe correlators
$\vec\phi^s(t)$.

More formally, we introduce $\sigma$-independent scaling correlators $F$
and $\vec F^s$ by
\begin{align}
  F(\bar t)&:=\lim_{\varepsilon\to0^-}\phi(t)\,,&
  \vec F^s_\delta(\tilde t)&:=\lim_{\varepsilon\to0^-}\vec\phi^s(t)\,,
\end{align}
with reduced times $\bar t=t/t_\sigma'$ and $\tilde t=t/t_{\sigma,\delta}'$.
From Eqs.~\eqref{equmotlaplace}, we obtain after taking the scaling limit
$\varepsilon\to0$ with $t_\sigma',t_{\sigma,\delta}'\to\infty$:
\begin{subequations}\label{alphascale}
\begin{align}
  \hat F(\bar z)&=\hat{\mathcal F}^c[F](\bar z)
  \cdot(1+\bar z\hat F(\bar z))\,,\\
  \label{alphascaletagged}
  \hat{\vec F}^s_\delta(\tilde z)
  &=\vec{\mathcal C}^s_c\left[\widehat{F\vec F^s_\delta}(\tilde z),
    \tilde z\hat{\vec F}^s_\delta(\tilde z)+\vec\phi_0^s\right]\,.
\end{align}
\end{subequations}
Here, $\mathcal F[F]$ denotes
the memory kernel of the host-liquid model, evaluated with the scaling
correlator.
The initial conditions for Eqs.~\eqref{alphascale} are obtained
by matching them to the long-time limit of the $\beta$-correlation
regime in the liquid, i.e.\ the von~Schweidler law. Hence,
$F(0)=f_c$, and $\lim_{\bar t\to0}(F(\bar t)-f_c){\bar t}^{-b}=-B$. This
leads to the standard
$\alpha$-scaling regime discussed in detail earlier \cite{Goetze.2009}.
Likewise, we identify $\vec F_\delta^s(0)=\vec f^s_c$ and
$\lim_{\tilde t\to0}(\vec F_\delta^s(\tilde t)-\vec f^s_c){\tilde t}^{-b}=-B^s$
where $B^s$ is trivially related to $B$ via the critical amplitudes.
As the plateau value
$\vec f^s_c$ vanishes linearly with $\delta$ upon approaching the
delocalization transition, it is reasonable to
introduce the probe-particle $\alpha$-scaling correlator
\begin{subequations}\label{alphascaling}
\begin{equation}
  \vec F^s(\tilde t):=\lim_{\delta\to0^-}\vec F^s_\delta(\tilde t)/|\delta|\,,
\end{equation}
to which corresponds the $\alpha$-scaling law
\begin{equation}
  \lim_{\delta\to0^-}\lim_{\varepsilon\to0}\vec\phi^s(t)/|\delta|
  =\vec F^s(t/t_{\sigma,\delta}')\,.
\end{equation}
\end{subequations}

It remains to be shown that $\vec F^s(\tilde t)$ indeed is independent
on $\sigma$ and $\delta$. To see this, expand Eq.~\eqref{alphascaletagged},
\begin{align}\label{alphares}
  0&=\A^{s,c}_c\cdot\tilde z\hat{\vec F}^s(\tilde z)
  \nonumber\\ &
  +|\delta|\left\{-\frac{d\A^{s,c}_c}{d\delta}\cdot
    \tilde z\hat{\vec F}^s(\tilde z)
  +f_c\vec{\mathcal C}^{s,c}_c\left[\tilde z\hat{\vec F}^s(\tilde z),
    \tilde z\hat{\vec F}^s(\tilde z)\right]
  \right.\nonumber\\ & \left.
  -B^s\vec{\mathcal C}^{s,c}_c\left[\tilde z\widehat{{\tilde t}^b
    \vec F^s}(\tilde z),\vec\phi_0^s\right]\right\}
  +\A^{s,c}_c\cdot\mathcal O(\delta)\,.
\end{align}
This suggests the ansatz
$\vec F^s(\tilde t)=F^s(\tilde t)\vec h^{s,c}_c+\mathcal O(\delta)$,
and multiplying Eq.~\eqref{alphares} with the left-null-eigenvector
$\hat{\vec h}^{s,c}_c$ yields an equation that is independent on $\delta$:
\begin{multline}
  0=-\tilde z\hat F^s(\tilde z)\hat{\vec h}^{s,c}_c\cdot
  \frac{d\A^{s,c}_c}{d\delta}\cdot\vec h^{s,c}_c\\
  +f_c(\tilde z\hat F^s(\tilde z))^2
  \hat{\vec h}^{s,c}_c\cdot\vec{\mathcal C}^{s,c}_c[\vec h^{s,c}_c,\vec h^{s,c}_c]
  \\
  -B^s\tilde z\widehat{{\tilde t}^bF^s}(\tilde z)
  \hat{\vec h}^{s,c}_c\cdot{\mathcal C}^{s,c}_c[\vec h^{s,c}_c,\vec\phi_0^s]\,.
\end{multline}
Consequently, $\vec F^s$ is invariant under rescaling of both variables
$\delta$ and $\varepsilon$. It depends, however,
on the direction of the path crossing the delocalization transition with
respect to the transition hypersurface.

\begin{figure}
\centerline{\includegraphics[width = \columnwidth]{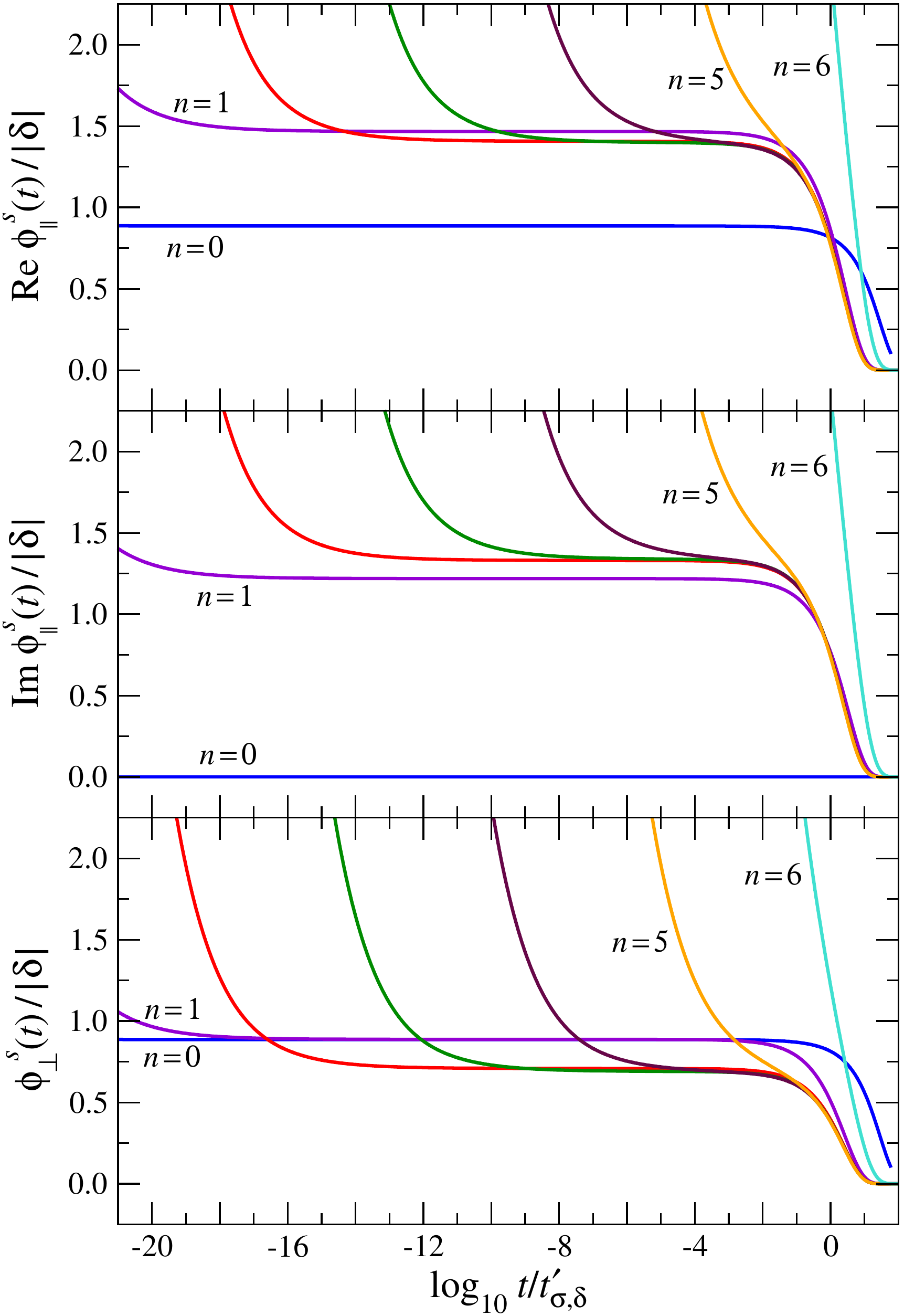}}
\caption{\label{fig:corr_alpha}
  Test of the $\alpha$-sclaing law for the probe-particle correlation
  functions: the $\vec\phi^s_\alpha(t)/|\delta|$ are shown as functions
  of rescaled time $\tilde t=t/t_{\sigma,\delta}'$. Parameters are chosen
  as in Fig.~\ref{fig:corr_overview}, with $\varepsilon=-10^{-12}$.
  Reduced forces are given by $\delta=-10^{-n}$, and $n=0$, $1$, $2$, $3$,
  $4$, $5$, and $6$ as labeled.
}
\end{figure}

The scaling property of the probe correlators as a function of $\delta\to0^-$
in the limit $\varepsilon\to0$ is exhibited by Fig.~\ref{fig:corr_alpha}.
Here, some very small $\varepsilon<0$ has been chosen, and the correlation
functions of the probe particle are plotted as
$\phi^s_\alpha(\tilde t)/|\delta|$
for various $\delta<0$. Reducing $|\delta|$, the $\alpha$-scaling regime
is entered, as seen for the $n=2$ through $n=5$ curves: as functions
of rescaled time $\tilde t=t/t_{\sigma,\delta}'$, all the correlators agree
with the scaling function $F^s_\alpha(\tilde t)$ at long times. The latter
shows an initial plateau, followed by a relaxation on a time scale
$\tilde t=\mathcal O(1)$. Qualitatively, this scaling is similar to the
usual $\alpha$-relaxation scaling law one finds without external force
upon varying $\varepsilon\to0$. There, however, the discontinuous nature
of the MCT transition predicts that asymptotically, the plateau seen in the
$\alpha$ correlator is constant, while for the delocalization transition
approached as $\delta\to0$, the plateau scales with the distance $|\delta|$
to the transition point.
For $n=0$ and $n=1$, the distance $|\delta|$ to
the delocalization transition is too large, and the scaling prediction
of Eq.~\eqref{alphascaling} no longer holds, as it requires both
$\delta\to0$ and $\varepsilon\to0$. The curve for $n=6$ also shown in
Fig.~\ref{fig:corr_alpha} corresponds to the regime
$|\delta^2/\sigma|\ll1$, and no longer satisfies the requirement
$\varepsilon\to0$ in Eq.~\eqref{alphascaling}; it hence deviates again
from the scaling law.

\section{The friction coefficient}\label{sec:zeta}

\subsection{Above the Delocalization Threshold}

We now focus on a discussion of the friction coefficient $\Delta\zeta$
in the asymptotic regime. Let us start by considering $\delta>0$, i.e.,
a force exceeding the delocalization threshold, so that even in the glass,
$\Delta\zeta<\infty$.

\begin{figure}
\centerline{\includegraphics[width = \columnwidth]{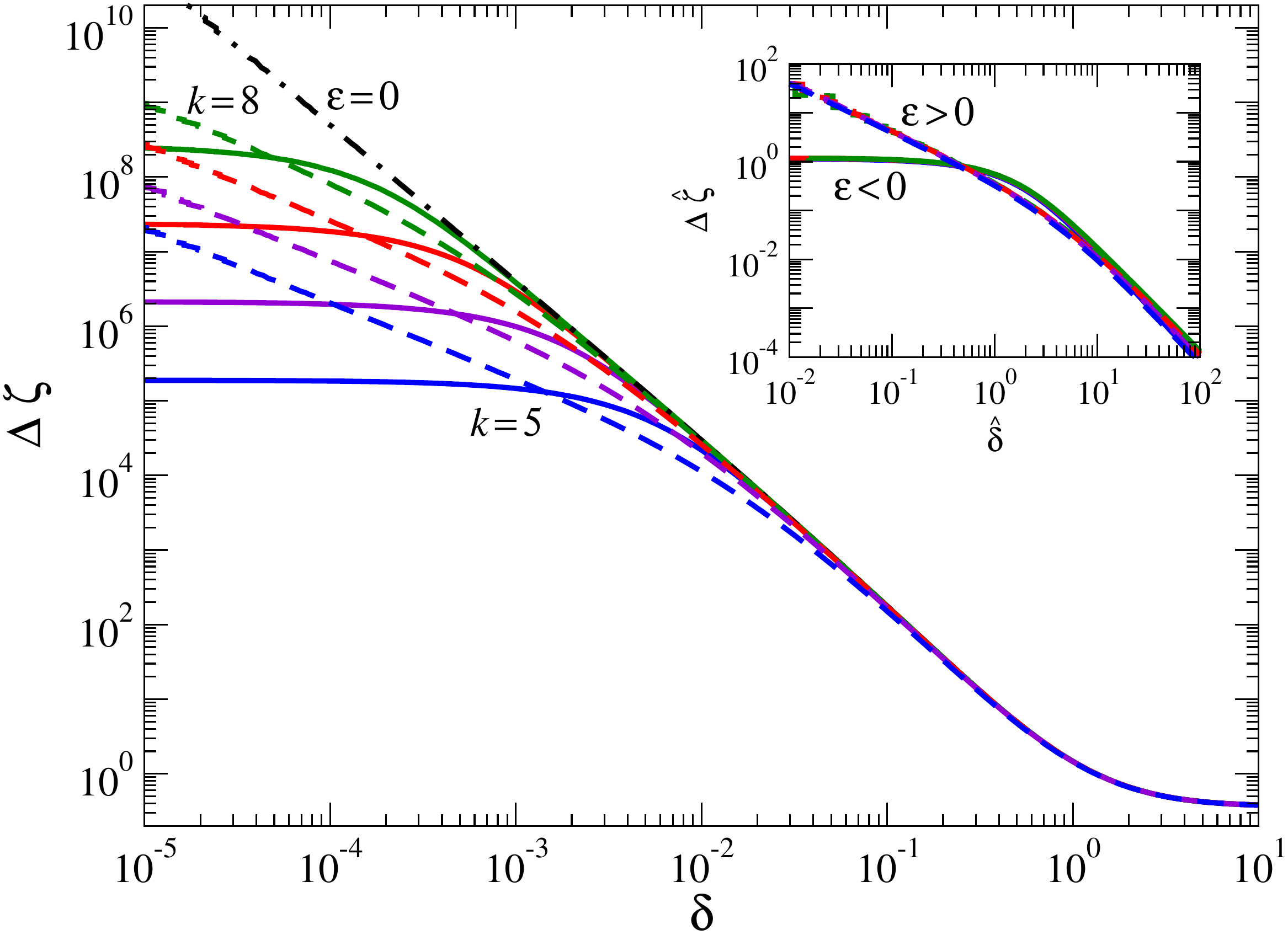}}
\caption{\label{fig:zeta_beta}
  Friction coefficient increment $\Delta\zeta=\zeta(\Fex)-\zeta_0$ for
  a probe particle pulled with external force $\Fex$, as a function of
  the distance to the delocalization threshold, $\delta=(\Fex-\Fexc)/\Fexc$,
  for $\delta>0$ and various distances to the glass transition $\varepsilon
  =\pm10^{-k}$ with $k=5$, $6$, $7$, and $8$ as labeled. Dashed lines are
  in the glass ($\varepsilon>0$), solid lines in the liquid ($\varepsilon<0$).
  The curve at the glass transition, $\varepsilon=0$, is shown as a
  dash-dotted line.
  Inset: curves for $\varepsilon\neq0$, as a scaling plot,
  $\Delta\hat\zeta=\Delta\zeta/|\varepsilon|^{1/2-1/(2a)}$ versus
  $\hat\delta=\delta/|\varepsilon|^{1/2}$.
}
\end{figure}

Figure~\ref{fig:zeta_beta} shows the friction coefficients from
Fig.~\ref{fig:zeta_overview} in a double-logarithmic representation
as a function of the distance to the delocalization threshold, $\delta>0$.
At large $\delta$, the qualitative features discussed before in
conjunction with Fig.~\ref{fig:zeta_overview} are recovered: a plateau
as $\delta\to\infty$, and an intermediate decay for $\delta\ll1$ that
we will identify with a power law. For $\delta\ll\sqrt{|\sigma|}$, this
power law crosses over: for liquid states, the curves approach a constant,
since the friction at the critical force ($\delta=0$) is finite there.
For glassy states, the friction diverges as $\delta\to0^+$, but with a
different power law than the one observed for $\delta\gg\sqrt{|\sigma|}$.

Consider first the regime dominated by the $\delta\to0$ limit: here,
Eq.~\eqref{decay12} describes the power-law decay of $\vec G^s(t)\sim t^{-1/2}$
for $\sigma>0$,
valid in a time window $t_\sigma\ll t\ll t_{1/2}$. Inserting into
Eq.~\eqref{delzeta}, one obtains
$
\Delta \zeta \approx \int_{\mathcal{O}(t_{\sigma})}^{t_{1/2}} C^{\prime} \left|\sigma\right|^{1/2 - 1/(4a)} \cdot \left(t_0/t\right)^{1/2} \cdot f
$
which results in
\begin{subequations}\label{dres1}
\begin{equation}
 \Delta \zeta (\delta)
   \propto \left|\sigma\right|^{1 - 1/(2a)} \cdot \delta^{-1}\,,
  \quad \text{$\sigma> 0$, $|\delta^2/\sigma| \ll 1$.}
\end{equation}
In the liquid state, the $t^{-1/2}$ power law does not appear. For
$\delta\sim\sqrt{|\sigma|}$ we get the $\delta$-independent
result in the liquid,
\begin{equation}
 \Delta \zeta (\delta)
  \propto \left|\sigma\right|^{1/2 - 1/(2a)} \,, 
  \quad \text{$\sigma< 0$, $|\delta^2/\sigma| \ll 1$.}
\end{equation}
\end{subequations}
Equations~\eqref{dres1} explain the qualitative behavior seen for $\delta\to0$
in Fig.~\ref{fig:zeta_beta}.

For $t_{\delta} \ll t_{\sigma}$, the probe $\beta$ correlator $\vec{G}^s(t)$
for times $t_0 \ll t \ll t_{\sigma}$ follows the power law $(t_0/t)^a$.
Integration of Eq.~\eqref{delzeta} leads to
\begin{equation}
  \Delta \zeta \propto \delta^{-(1-a)/a} \,,
  \quad\text{$|\delta^2/\sigma| \gg 1$,}
\end{equation}
where the proportionality constant is independent of $\sigma$.
This power law is seen in Fig.~\ref{fig:zeta_beta} in the intermediate-$\delta$
regime, where $\delta\gg\sqrt{|\sigma|}$ but still small.

Having identified these power laws, a scaling prediction is obtained for
the friction coefficient, for forces just above the threshold. Introducing
$\hat\delta=\delta/\sqrt{|\varepsilon|}$ and
$\Delta\hat\zeta=\Delta\zeta/|\varepsilon|^{1/2-1/(2a)}$, we obtain two
master curves, one for $\varepsilon>0$, and one for $\varepsilon<0$. This
is shown in the inset of Fig.~\ref{fig:zeta_beta}.
The scaling curve for the glass shows two power laws,
$\Delta\hat\zeta\sim\hat\delta^{-1}$ for $\hat\delta\ll1$,
and $\Delta\hat\zeta\sim\hat\delta^{1-1/a}$ for $\hat\delta\gg1$. For
our model, $a\approx1/3$, so that the exponent governing the large-$\hat\delta$
decay in $\Delta\hat\zeta$ is close to 2, as seen by inspection of
Fig.~\ref{fig:zeta_beta}.
One has to note that the identification of these power laws in experiment
or simulation is hampered by the fact that one has to approach both
$\varepsilon\to0$ and $\delta\to0$. For finite distances to the transition,
preasymptotic corrections quickly become dominant. Fits using the schematic
model we discuss here have to be performed with varying coupling coefficient
$v^s$ \cite{Gnann.2011}. Compared to the case displayed by
Fig.~\ref{fig:zeta_beta}, this introduces a regular shift of the curves
\cite{Gnann.2009}.

\subsection{Below the Delocalization Threshold}

We now discuss the case $\delta<0$. Here, the probe remains localized within
glassy states, so that for $\delta<0$ and $\sigma>0$,
$\Delta\zeta=\infty$ holds trivially. For $\sigma<0$, the final
$\alpha$ relaxation will give the dominant contribution to the friction
coefficient, which is obtained by inserting the $\alpha$-scaling law
into Eq.~\eqref{delzeta},
\begin{multline}
 \Delta \zeta \approx (1-\mu) \int_0^{\infty} F(t/t_\sigma')
  F_{\delta,1}^s(t/t_{\sigma,\delta}')\, dt\\
 + \mu \int_0^{\infty} F(t/t_\sigma')
  F_{\delta,3}^s(t/t_{\sigma,\delta}')\, dt\,.
\end{multline}
This expression yields a factorization of $\Delta \zeta$ of the form
$
  \Delta\zeta\sim|\sigma|^{-1/(2a)-1/(2b)}Z(\delta)
$
for $\sigma\to0$,
where $Z$ is universal. If $t_{\sigma,\delta}' \gg t_\sigma$, i.e.,
$|\delta^2/\sigma|\gg1$, and additionally $|\delta|\ll1$,
we further use that $\vec F_\delta^s(t/t_{\sigma,\delta}')$ is
proportional to $|\delta|$. From the von~Schweidler law we then arrive at
\begin{equation}\label{dres2}
 \Delta \zeta \propto |\sigma|^{-1/(2a)-1/(2b)} |\delta|^{1/b+1}\,,
  \quad\text{$\sqrt{|\sigma|}\ll|\delta|\ll1$.}
\end{equation}

\begin{figure}
\centerline{\includegraphics[width = \columnwidth]{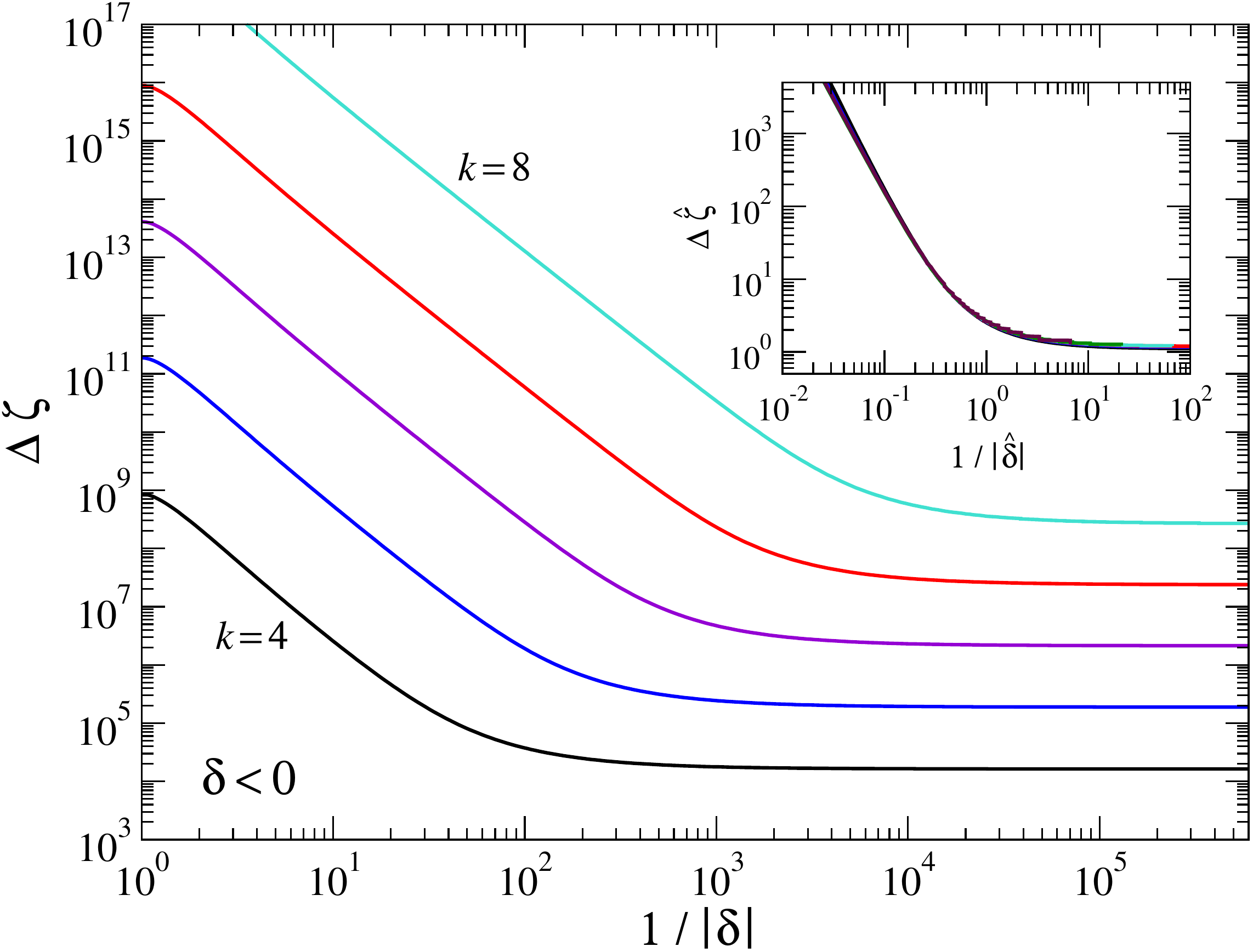}}
\caption{\label{fig:zeta_alpha}
  Friction coefficient increment $\Delta\zeta$ as a function of the distance
  to the delocalization threshold $\delta$, for $\delta<0$ (i.e.,
  $\Fex<\Fexc$) in the liquid. The distance to the glass transition is
  chosen as $\varepsilon=-10^{-k}$, with $k=4$, $5$, $6$, $7$, and $8$ as
  labeled.
  Inset: scaling plot $\Delta\hat\zeta$ versus $\hat\delta$ for the same
  data and also including $k=9$ and $k=10$.
}
\end{figure}

Hence, the exponent of von Schweidler's law is also present in the behavior
of $\Delta \zeta$, as can bee seen in Fig.~\ref{fig:zeta_alpha}, where
$\Delta(\zeta)$ is plotted as a function of $1/|\delta|$ for forces
below the delocalizaton threshold, in the liquid.
Note that for $\Fex\to0$, i.e., $\delta\to-1$, the von~Schweidler regime
is cut off due to the microscopic relaxation of the correlation functions,
and $\Delta\zeta(\Fex\!=\!0)-\Delta\zeta\sim\Fex^2$ is obtained as mentioned
further above.
As the glass transition
is approached, this window of initial linear response shrinks.

Again, Eq.~\eqref{dres2} includes a scaling prediction: plotting
$\Delta\hat\zeta$ as a function of $\hat\delta$, where the rescaled
variables are defined above, the curves for all $\varepsilon$ close to
the glass transition can be scaled onto one master curve. This is shown
in the inset of Fig.~\ref{fig:zeta_alpha}. The von~Schweidler law is seen
for $\hat\delta\ll1$, in the form $\Delta\hat\zeta\sim\hat\delta^{1/b+1}$.
This exponent is close to $2.6$ for the model parameters chosen here, as
one can verify in the figure.

\subsection{Large-Force Plateau}

We come briefly back to the case, where the probe particle is delocalized.
For $\delta \gg \mathcal{O}(1)$, the $\beta$-scaling regime does not give the
dominant contribution to $\Delta \zeta$ any longer and the short-time dynamics
of the correlation functions is more important. In this case it is sufficient
to consider the probe dynamics alone and that of the host liquid as essentially
arrested, i.e.,
$v_j^s\phi(t)\approx v_{j,\text{eff}}^s$ as constants. The friction
coefficient for $\delta\to\infty$, $\Delta\zeta_\infty=\Delta\zeta(\Fex\to\infty)$ can then be calculated as
\begin{multline}
  \Delta\zeta_\infty\approx\phi_0\int_0^\infty dt\,\left\{
  (1-\mu)\Real\phi^s_\parallel(t)
  +\mu\phi^s_\perp(t)\right\}\\
  =\phi_0\left\{(1-\mu)\Imag\hat\phi^s_1(z\!=\!0)
  +\mu\Imag\hat\phi^s_3(z\!=\!0)\right\}\,.
\end{multline}
From Eq.~\eqref{sjoephilt}, $\vec\psi^s=\hat{\vec\phi}^s(z\!=\!0)$
fulfills a linear equation system,
$
  \cdot\vec\psi^s = i\tensor\omega^{-1}\vec\phi_0^s+\vec{\mathcal C}^s[
  \vec\psi^s,\vec\phi_0^s]
$.
This yields an algebraic approximation for $\Delta\zeta_\infty$.
For the model discussed here, we get
\begin{equation}
  \Delta\zeta_\infty\sim\mu f_\text{eff}+\mathcal O(1/\Fex^2)\,,
\end{equation}
where $f_\text{eff}=\mathcal O(1)$ is the approximately constant value
of the host-liquid correlator $\phi(t)$ for the time window over which
the probe-particle correlations decay.
The increment of the friction coefficient as $\Fex\to\infty$
hence is given by the parameter
$\mu$ that controls the admixture of ``perpendicular'' modes to the
integral determining $\Delta\zeta$. This fact has been exploited
in Ref.~\cite{Gnann.2011} in order to quantitatively describe the large-force
amplitude of measured data. Setting $\mu=0$ results in an
expression for $\Delta\zeta$ that decays to zero as $1/\Fex^2$ at large
forces. Setting further $v_2^s=0$, one obtains the model used in
Ref.~\cite{Gazuz.2009}, where
$\Delta\zeta_\infty=f_\text{eff}(1+v^s_\text{eff})
/(1-{v_\text{eff}^s}^2+\Fex^2)$. This expression holds for $v^s_\text{eff}\ge1$,
ensuring positivity of the result.

\begin{figure}
\centerline{\includegraphics[width = \columnwidth]{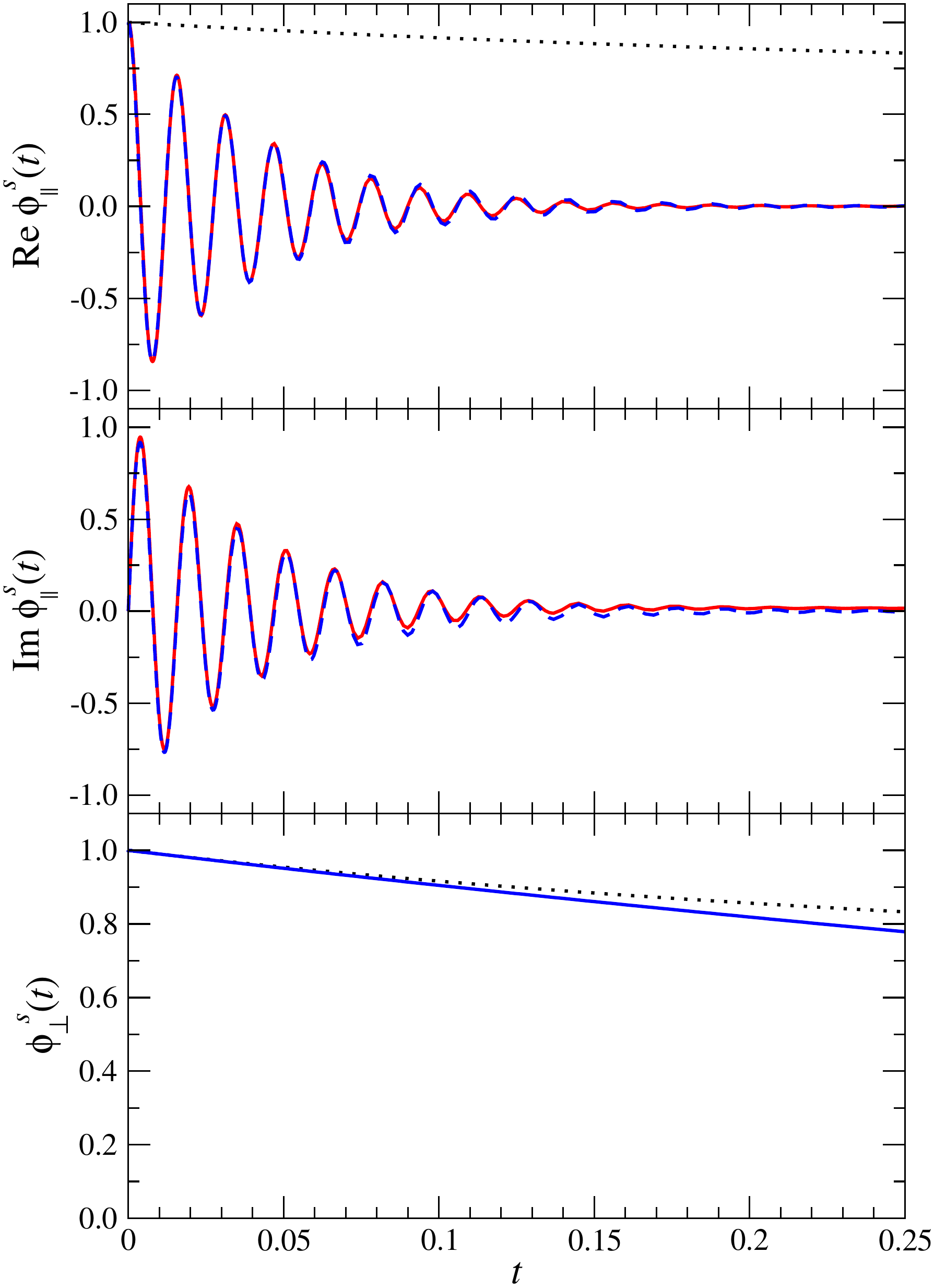}}
\caption{\label{fig:corr_os}
  Correlations functions $\phi^s_\parallel(t)$ and $\phi^s_\perp(t)$ of
  the schematic model, with parameters as in Fig.~\ref{fig:corr_overview},
  but a large external force, $\Fex=402.7$.
  Full lines show the numerical solutions of the model, dashed lines the
  least-square fits using an exponentially damped function
  $\phi^s_\parallel(t)=e^{-\nu t}e^{i(\Fex+\xi)t}$ with parameters
  $\nu=22.6$ and $\xi=-2.1$.
  The host-liquid correlator $\phi(t)$ is shown as a dotted line in
  the top and bottom panels.
}
\end{figure}

The reason that the correlation function $\phi_\parallel^s(t)$ does not
contribute to the friction-coefficient increment at large forces, is its
oscillatory decay. Fig.~\ref{fig:corr_os} shows exemplary results for
a very large force $\Fex$: both $\phi(t)$, shown as dotted lines, and
$\phi^s_\perp(t)$ decay roughly exponentially, with a relaxation time that
is much larger than the one relevant for $\phi^s_\parallel(t)$ in this
regime. The latter becomes proportional to a damped oscillation,
$\phi^s_\parallel(t)\approx\exp[-t+i\kappa_\parallel\Fex t]$ in the absence of any
memory-kernel damping, cf.\ Eq.~\eqref{sjoephi}. Taking into account the
remaining damping from the memory kernel, we find that $\phi^s_\parallel(t)$
for large $\Fex$ is nicely fitted with $\exp[-\nu t]\exp[i(\kappa_\parallel\Fex+\xi)t]$.
Such a fit is included in Fig.~\ref{fig:corr_os}, with parameters
$\nu=22.6$ and $\xi=-2.1$ determined by a least-square error minimization;
it is virtually indistinguishable from the numerical result for
$\phi^s_\parallel(t)$ in the figure.

The microscopic interpretation of these oscillations is a steady motion
of the probe, $\vec r^s(t)=\vec r^s(0)+\vec v\,t$, leading to
$\phi_{\vec q}^s(t)=\exp[i\vec q\cdot\vec v\,t]$. For $\vec q\parallel\vec\Fex$,
choosing a wave vector characterizing typical probe--host interaction
length scales, we recover $\zeta(\Fex\!\to\!\infty)=q_\parallel/\kappa_\parallel$.
The schematic-model parameter $\kappa_\parallel$ therefore relates typical length
scales to the high-force friction coefficient.
It is also clearly seen that for $\vec q\perp\vec\Fex$, no oscillations remain
in the correlation function; this is correctly captured in the schematic
model.

\section{Conclusion}\label{conclusion}

We have discussed a schematic mode-coupling model for the
nonlinear response of the friction coefficient in force-driven
active microrheology. Expressions describing the asymptotic behavior
of the probe-particle density correlation function in the directions
parallel and perpendicular to the force within the schematic model have been
derived. From these, together with the known asymptotic behavior of the
equilibrium host-liquid correlation function, we have inferred
two-parameter scaling laws that yield scaling forms for the
probe-particle friction coefficient. The two small parameters are
the distance to the glass transition of the host liquid, $\varepsilon$,
and the distance $\delta$ to the delocalization transition, where a frozen-in
probe becomes mobile in response to a force-induced local melting of the
host.
There exists a scaling limit based on the combination $|\varepsilon/\delta^2|$:
if this parameter is small, the external force is a small perturbation to
the glassy dynamics of the host liquid, while for
$|\delta|\ll\sqrt{|\varepsilon|}$, the vicinity to the force-induced
local melting of the glass dominates the dynamics.

The asymptotic expressions for the correlation functions lead to
the identification of five relevant time scales:
two scales, $t_\sigma$ and $t_\sigma'$, determine the power-law regimes
for the host-liquid correlator. These diverge upon approaching the glass
transition, $\sigma\to0$, and are independent of the external force
characterized by $\delta$. For the coupling of the probe correlators to
that of the host liquid, a time scale $t_\delta$ was identified that
diverges as the critical force is approached, $\delta\to0$. It separates
two regimes of proportionality, $\vec G^s(t)\propto G(t)$, but with different
prefactors. Considering the long-time decay of the probe correlator,
we identified two further time scales, $t_{1/2}$ and $t_{\sigma,\delta}'$,
describing, respectively, the decay inside a glassy host liquid and
the coupling of the probe relaxation to the host's von~Schweidler law.
For the friction coefficient, the various asymptotic regimes of the
correlation function induce several power-law regimes around the
critical threshold force, $\Fexc$, marking the point where the external
force becomes more effective in breaking cages than thermal fluctuations.

In the linear response regime, the well-known asymptotic
result of equilibrium MCT is recovered, stating that the structural relaxation
of the probe is, up to a wave-vector dependent amplitude, given by the
relaxation of the host, cf.\ Eq.~\eqref{betacoupling}.
In a sense, this is the justification of the discussion of microrheology
response in terms of generalized Stokes-Einstein relations, where one tries
to relate the diffusivity of the probe particle to the collective dynamics
of the surrounding host (expressed by its shear viscosity). Within MCT,
a relation of the form $D\sim1/\eta$ is the consequence of
Eq.~\eqref{betacoupling}, with prefactors that are typically of the size
expected from the Stokes-Einstein relation \cite{FuchsSE}.
Note, however, that the true Stokes-Einstein relation is of the form
$D\sim\kT/\eta$, where the factor $\kT$ identifies it as a relation
connecting a macroscopic transport coefficient to microscopic fluctuations.
Indeed, recent experiments indicate that the factor $\kT$ is absent in the
empirical relation between $D$ in $\eta$ found in viscous liquids
\cite{MeyerSE}, in agreement with Eq.~\eqref{betacoupling}.

Small external forces will preserve this coupling, but as the critical
force is approached, the range of validity of Eq.~\eqref{betacoupling}
continuously shrinks.
The initial deviation from the linear reponse regime, by symmetry, has
to be quadratic, $\Delta\zeta(\Fex)-\Delta\zeta(\Fex\!=\!0)\sim-\Fex^2$.
Approaching $\Fexc$, however, the von~Schweidler law governs the
force thinning behavior, i.e., a power law involving the exponent $b$
is seen in $\Delta\zeta$. This is, however, a small effect, not immediately
apparent from experimental data.

The delocalization transition is, within the present schematic model,
identified as a continuous transition, i.e. the nonergodicity parameter
of the probe-particle density correlators vanishes continuously. In
a microscopic model, this would correspond to a continuous broadening
of the probability distribution function for the probe position. This
can be understood as a single-particle localization length that diverges
due to the applied force. However, in the schematic model, no such
interpretation is obvious, since the model does not carry any information
related to large length scales. It is also expected that the
exponents found in the asymptotic analysis carried out here will be modified
by taking into account long-wavelength fluctuations
\cite{Schnyder.2011}.

Above the threshold force, tracer motion is always delocalized. For this
case, the friction coefficient increment $\Delta\zeta$ follows two power
laws: a trivial force thinning with exponent $-1$ is predicted in the glass,
valid for very small $|\Fex-\Fexc|$. This law is,
closer to the glass transition, replaced by a power law involving
the MCT critical exponent, i.e., force thinning with exponent
$y=1-1/a<-1$. Inserting typical values found for hard-sphere like
systems within MCT, $a\approx1/3$, one gets for the force thinning exponent
$y\approx-2$.

Considering still larger forces, $\Fex\to\infty$, a feature seen in
experiment is a second plateau in $\Delta\zeta>0$, indicating that the
friction experienced by the pulled particle is not just the solvent
friction. Although hydrodynamic interactions (HI) in the colloidal suspension
will play a major role in this regime \cite{Khair.2006},
the second plateau is not only due to them, as it is also found in
Brownian-dynamics simulations \cite{Gazuz.2009} that do not include HI.
Within the schematic model, the additional contribution can be traced
back to the correlation function $\phi_\perp^s(t)$ mimicking the
density correlators in the direction perpendicular to the force.

Although the schematic model we have discussed gives a reasonable
quantiative description of the friction coefficients, and a qualitative one
for intermediate-length-scale density correlation functions,
the question remains whether the force-driven delocalization
of a probe particle in the glass is continuous in the sense mentioned above.
Another possibility is that of a discontinuous local yielding, where
the plateau of the tagged-particle correlation function $\vec f^s$ does
not decrease to zero as $\Fex\to\Fexc-0$, but to some nonzero constant
(so that a discontinuous jump in the nonergodicity parameter at $\Fexc$
results). These scenarios are not easily distinguished following the behavior
of the friction coefficient. Correlation functions from computer simulation
of Brownian soft-sphere systems \cite{Gazuz.2009,Gnann.2011} appear
compatible with the continuous scenario incorporated in the schematic model
discussed here. However, recent MD simulations of the mean-squared displacement
in active microrheology show a plateau that does not change appreciably
with varying $\Fex$ \cite{Winter.2012}, and thus may point to a discontinuous
transition scenario. Note that even the connection between schematic and
``full'' microscopic MCT remains unclear at the moment. This connection
hinges upon the assumption that the two modes chosen for the probe correlator
in the schematic model are sufficient to represent the full set of critical
modes in the bifurcation transition of the full model. In this respect,
the microscopic model of Ref.~\cite{Gazuz.2009} is an extension of MCT that
is qualitatively different from other extensions of the original MCT where
it could be shown that the latter's bifurcation class $\mathcal A_l$ was
not left.




\begin{acknowledgments}
We thank M.~Fuchs and C.~Harrer for discussions.
This work was supported by DFG SFB-Transregio TR6, project A7.
Th.V.\ thanks for funding through the Helmholtz-Gemeinschaft
(HGF VH-NG~406), and the Zukunftskolleg der Universit\"at Konstanz.
M.V.G.\ thanks for funding through the research initiative Analysis
and Numerics of Evolution Equations with Applications in the Sciences
of the Universit\"at Konstanz and the graduate school IMPRS of the
Max Planck Institute for Mathematics in the Sciences. 
\end{acknowledgments}

\bibliography{microrheo_pub}

\begin{thebibliography}{35}
\expandafter\ifx\csname natexlab\endcsname\relax\def\natexlab#1{#1}\fi
\expandafter\ifx\csname bibnamefont\endcsname\relax
  \def\bibnamefont#1{#1}\fi
\expandafter\ifx\csname bibfnamefont\endcsname\relax
  \def\bibfnamefont#1{#1}\fi
\expandafter\ifx\csname citenamefont\endcsname\relax
  \def\citenamefont#1{#1}\fi
\expandafter\ifx\csname url\endcsname\relax
  \def\url#1{\texttt{#1}}\fi
\expandafter\ifx\csname urlprefix\endcsname\relax\def\urlprefix{URL }\fi
\providecommand{\bibinfo}[2]{#2}
\providecommand{\eprint}[2][]{\url{#2}}

\bibitem[{\citenamefont{Waigh}(2005)}]{Waigh.2005}
\bibinfo{author}{\bibfnamefont{T.~A.} \bibnamefont{Waigh}},
  \bibinfo{journal}{Rep. Prog. Phys.} \textbf{\bibinfo{volume}{68}},
  \bibinfo{pages}{685} (\bibinfo{year}{2005}).

\bibitem[{\citenamefont{Squires}(2008)}]{Squires.2008}
\bibinfo{author}{\bibfnamefont{T.~M.} \bibnamefont{Squires}},
  \bibinfo{journal}{Langmuir} \textbf{\bibinfo{volume}{24}},
  \bibinfo{pages}{1147} (\bibinfo{year}{2008}).

\bibitem[{\citenamefont{Erbe et~al.}(2008)\citenamefont{Erbe, Zientara,
  Baraban, Kreidler, and Leiderer}}]{Erbe.2008}
\bibinfo{author}{\bibfnamefont{A.}~\bibnamefont{Erbe}},
  \bibinfo{author}{\bibfnamefont{M.}~\bibnamefont{Zientara}},
  \bibinfo{author}{\bibfnamefont{L.}~\bibnamefont{Baraban}},
  \bibinfo{author}{\bibfnamefont{C.}~\bibnamefont{Kreidler}}, \bibnamefont{and}
  \bibinfo{author}{\bibfnamefont{P.}~\bibnamefont{Leiderer}},
  \bibinfo{journal}{J. Phys.: Condens. Matt.} \textbf{\bibinfo{volume}{20}},
  \bibinfo{pages}{404215} (\bibinfo{year}{2008}).

\bibitem[{\citenamefont{Wilhelm}(2008)}]{Wilhelm.2008}
\bibinfo{author}{\bibfnamefont{C.}~\bibnamefont{Wilhelm}},
  \bibinfo{journal}{Phys. Rev. Lett.} \textbf{\bibinfo{volume}{101}},
  \bibinfo{pages}{028101} (\bibinfo{year}{2008}).

\bibitem[{\citenamefont{Hastings et~al.}(2003)\citenamefont{Hastings, {Olson
  Reichhardt}, and Reichhardt}}]{Hastings.2003}
\bibinfo{author}{\bibfnamefont{M.~B.} \bibnamefont{Hastings}},
  \bibinfo{author}{\bibfnamefont{C.~J.} \bibnamefont{{Olson Reichhardt}}},
  \bibnamefont{and}
  \bibinfo{author}{\bibfnamefont{C.}~\bibnamefont{Reichhardt}},
  \bibinfo{journal}{Phys. Rev. Lett.} \textbf{\bibinfo{volume}{90}},
  \bibinfo{pages}{098302} (\bibinfo{year}{2003}).

\bibitem[{\citenamefont{Habdas et~al.}(2004)\citenamefont{Habdas, Schaar,
  Levitt, and Weeks}}]{Habdas.2004}
\bibinfo{author}{\bibfnamefont{P.}~\bibnamefont{Habdas}},
  \bibinfo{author}{\bibfnamefont{D.}~\bibnamefont{Schaar}},
  \bibinfo{author}{\bibfnamefont{A.~C.} \bibnamefont{Levitt}},
  \bibnamefont{and} \bibinfo{author}{\bibfnamefont{E.~R.} \bibnamefont{Weeks}},
  \bibinfo{journal}{Europhys. Lett.} \textbf{\bibinfo{volume}{67}},
  \bibinfo{pages}{477} (\bibinfo{year}{2004}).

\bibitem[{\citenamefont{Williams and Evans}(2006)}]{Williams.2006}
\bibinfo{author}{\bibfnamefont{S.~R.} \bibnamefont{Williams}} \bibnamefont{and}
  \bibinfo{author}{\bibfnamefont{D.~J.} \bibnamefont{Evans}},
  \bibinfo{journal}{Phys. Rev. Lett.} \textbf{\bibinfo{volume}{96}},
  \bibinfo{pages}{015701} (\bibinfo{year}{2006}).

\bibitem[{\citenamefont{Squires and Mason}(2010)}]{Squires.2010}
\bibinfo{author}{\bibfnamefont{T.~M.} \bibnamefont{Squires}} \bibnamefont{and}
  \bibinfo{author}{\bibfnamefont{T.~G.} \bibnamefont{Mason}},
  \bibinfo{journal}{Annu. Rev. Fluid Mech.} \textbf{\bibinfo{volume}{42}},
  \bibinfo{pages}{413} (\bibinfo{year}{2010}).

\bibitem[{\citenamefont{Gazuz et~al.}(2009)\citenamefont{Gazuz, Puertas,
  Voigtmann, and Fuchs}}]{Gazuz.2009}
\bibinfo{author}{\bibfnamefont{I.}~\bibnamefont{Gazuz}},
  \bibinfo{author}{\bibfnamefont{A.~M.} \bibnamefont{Puertas}},
  \bibinfo{author}{\bibfnamefont{{\relax Th}.}~\bibnamefont{Voigtmann}},
  \bibnamefont{and} \bibinfo{author}{\bibfnamefont{M.}~\bibnamefont{Fuchs}},
  \bibinfo{journal}{Phys. Rev. Lett.} \textbf{\bibinfo{volume}{102}},
  \bibinfo{pages}{248302} (\bibinfo{year}{2009}).

\bibitem[{\citenamefont{Gnann et~al.}(2011)\citenamefont{Gnann, Gazuz, Puertas,
  Fuchs, and Voigtmann}}]{Gnann.2011}
\bibinfo{author}{\bibfnamefont{M.~V.} \bibnamefont{Gnann}},
  \bibinfo{author}{\bibfnamefont{I.}~\bibnamefont{Gazuz}},
  \bibinfo{author}{\bibfnamefont{A.~M.} \bibnamefont{Puertas}},
  \bibinfo{author}{\bibfnamefont{M.}~\bibnamefont{Fuchs}}, \bibnamefont{and}
  \bibinfo{author}{\bibfnamefont{{\relax Th}.}~\bibnamefont{Voigtmann}},
  \bibinfo{journal}{Soft Matter} \textbf{\bibinfo{volume}{7}},
  \bibinfo{pages}{1390} (\bibinfo{year}{2011}).

\bibitem[{\citenamefont{G\"otze and Sj\"ogren}(1987)}]{Goetze.1987}
\bibinfo{author}{\bibfnamefont{W.}~\bibnamefont{G\"otze}} \bibnamefont{and}
  \bibinfo{author}{\bibfnamefont{L.}~\bibnamefont{Sj\"ogren}},
  \bibinfo{journal}{Z. Phys. B} \textbf{\bibinfo{volume}{65}},
  \bibinfo{pages}{415} (\bibinfo{year}{1987}).

\bibitem[{\citenamefont{Fuchs and Cates}(2003)}]{Fuchs.2003}
\bibinfo{author}{\bibfnamefont{M.}~\bibnamefont{Fuchs}} \bibnamefont{and}
  \bibinfo{author}{\bibfnamefont{M.~E.} \bibnamefont{Cates}},
  \bibinfo{journal}{Faraday Discuss.} \textbf{\bibinfo{volume}{123}},
  \bibinfo{pages}{267} (\bibinfo{year}{2003}).

\bibitem[{\citenamefont{G\"otze and L.~Sj\"ogren}(1995)}]{Goetze.1995}
\bibinfo{author}{\bibfnamefont{W.}~\bibnamefont{G\"otze}} \bibnamefont{and}
  \bibinfo{author}{\bibfnamefont{L.}~\bibnamefont{L.~Sj\"ogren}},
  \bibinfo{journal}{J. Math. Analysis Appl.} \textbf{\bibinfo{volume}{195}},
  \bibinfo{pages}{230} (\bibinfo{year}{1995}).

\bibitem[{\citenamefont{Franosch and Voigtmann}(2002)}]{Franosch.2002}
\bibinfo{author}{\bibfnamefont{T.}~\bibnamefont{Franosch}} \bibnamefont{and}
  \bibinfo{author}{\bibfnamefont{{\relax Th}.}~\bibnamefont{Voigtmann}},
  \bibinfo{journal}{J.~Stat.\ Phys.} \textbf{\bibinfo{volume}{109}},
  \bibinfo{pages}{237} (\bibinfo{year}{2002}).

\bibitem[{\citenamefont{G\"otze}(2009)}]{Goetze.2009}
\bibinfo{author}{\bibfnamefont{W.}~\bibnamefont{G\"otze}},
  \emph{\bibinfo{title}{Complex Dynamics of Glass-Forming Liquids}}
  (\bibinfo{publisher}{Oxford University Press}, \bibinfo{year}{2009}).

\bibitem[{\citenamefont{Fuchs and Cates}(2009)}]{Fuchs.2009}
\bibinfo{author}{\bibfnamefont{M.}~\bibnamefont{Fuchs}} \bibnamefont{and}
  \bibinfo{author}{\bibfnamefont{M.~E.} \bibnamefont{Cates}},
  \bibinfo{journal}{J. Rheol. (NY)} \textbf{\bibinfo{volume}{53}},
  \bibinfo{pages}{957} (\bibinfo{year}{2009}).

\bibitem[{\citenamefont{Hajnal and Fuchs}(2009)}]{Hajnal.2009}
\bibinfo{author}{\bibfnamefont{D.}~\bibnamefont{Hajnal}} \bibnamefont{and}
  \bibinfo{author}{\bibfnamefont{M.}~\bibnamefont{Fuchs}},
  \bibinfo{journal}{Eur. Phys. J. E} \textbf{\bibinfo{volume}{28}},
  \bibinfo{pages}{125} (\bibinfo{year}{2009}).

\bibitem[{\citenamefont{Kr\"uger et~al.}(2011)\citenamefont{Kr\"uger, Weysser,
  and Fuchs}}]{Krueger.2011}
\bibinfo{author}{\bibfnamefont{M.}~\bibnamefont{Kr\"uger}},
  \bibinfo{author}{\bibfnamefont{F.}~\bibnamefont{Weysser}}, \bibnamefont{and}
  \bibinfo{author}{\bibfnamefont{M.}~\bibnamefont{Fuchs}},
  \bibinfo{journal}{Eur. Phys. J. E} \textbf{\bibinfo{volume}{34}},
  \bibinfo{pages}{88} (\bibinfo{year}{2011}).

\bibitem[{\citenamefont{Fuchs and Cates}(2002)}]{Fuchs.2002}
\bibinfo{author}{\bibfnamefont{M.}~\bibnamefont{Fuchs}} \bibnamefont{and}
  \bibinfo{author}{\bibfnamefont{M.~E.} \bibnamefont{Cates}},
  \bibinfo{journal}{Phys. Rev. Lett.} \textbf{\bibinfo{volume}{89}},
  \bibinfo{pages}{248304} (\bibinfo{year}{2002}).

\bibitem[{\citenamefont{Sj\"ogren}(1986)}]{Sjoegren.1986}
\bibinfo{author}{\bibfnamefont{L.}~\bibnamefont{Sj\"ogren}},
  \bibinfo{journal}{Phys. Rev. A} \textbf{\bibinfo{volume}{33}},
  \bibinfo{pages}{1254} (\bibinfo{year}{1986}).

\bibitem[{\citenamefont{N\"agele}(1996)}]{Naegele.1996}
\bibinfo{author}{\bibfnamefont{G.}~\bibnamefont{N\"agele}},
  \bibinfo{journal}{Phys. Rep.} \textbf{\bibinfo{volume}{272}},
  \bibinfo{pages}{215} (\bibinfo{year}{1996}).

\bibitem[{\citenamefont{Voigtmann}(2011)}]{Voigtmann.2011b}
\bibinfo{author}{\bibfnamefont{{\relax Th}.}~\bibnamefont{Voigtmann}},
  \bibinfo{journal}{EPL} \textbf{\bibinfo{volume}{96}}, \bibinfo{pages}{36006}
  (\bibinfo{year}{2011}).

\bibitem[{\citenamefont{Squires and Brady}(2005)}]{Squires.2005}
\bibinfo{author}{\bibfnamefont{T.~M.} \bibnamefont{Squires}} \bibnamefont{and}
  \bibinfo{author}{\bibfnamefont{J.~F.} \bibnamefont{Brady}},
  \bibinfo{journal}{Phys. Fluids} \textbf{\bibinfo{volume}{17}},
  \bibinfo{pages}{073101} (\bibinfo{year}{2005}).

\bibitem[{\citenamefont{Carpen and Brady}(2005)}]{Carpen.2005}
\bibinfo{author}{\bibfnamefont{I.~C.} \bibnamefont{Carpen}} \bibnamefont{and}
  \bibinfo{author}{\bibfnamefont{J.~F.} \bibnamefont{Brady}},
  \bibinfo{journal}{J. Rheol.} \textbf{\bibinfo{volume}{49}},
  \bibinfo{pages}{1483} (\bibinfo{year}{2005}).

\bibitem[{\citenamefont{Widder}(1972)}]{Widder.1972}
\bibinfo{author}{\bibfnamefont{D.}~\bibnamefont{Widder}},
  \emph{\bibinfo{title}{The Laplace Transform}} (\bibinfo{publisher}{Princeton
  University Press}, \bibinfo{year}{1972}), \bibinfo{edition}{8th} ed.

\bibitem[{\citenamefont{Arnol'd}(1992)}]{Arnold.1992}
\bibinfo{author}{\bibfnamefont{V.}~\bibnamefont{Arnol'd}},
  \emph{\bibinfo{title}{Catastrophe theory}} (\bibinfo{publisher}{Springer},
  \bibinfo{address}{Berlin}, \bibinfo{year}{1992}), \bibinfo{edition}{3rd} ed.

\bibitem[{\citenamefont{G{\"o}tze}(1991)}]{Goetze1991a}
\bibinfo{author}{\bibfnamefont{W.}~\bibnamefont{G{\"o}tze}}, in
  \emph{\bibinfo{booktitle}{Liquids, Freezing and Glass Transition (Les
  Houches, Session LI)}}, edited by \bibinfo{editor}{\bibfnamefont{J.~P.}
  \bibnamefont{Hansen}},
  \bibinfo{editor}{\bibfnamefont{D.}~\bibnamefont{Levesque}}, \bibnamefont{and}
  \bibinfo{editor}{\bibfnamefont{J.}~\bibnamefont{Zinn-Justin}}
  (\bibinfo{publisher}{North-Holland}, \bibinfo{address}{Amsterdam, Oxford, New
  York}, \bibinfo{year}{1991}).

\bibitem[{\citenamefont{G\"otze}(1990)}]{Goetze.1990}
\bibinfo{author}{\bibfnamefont{W.}~\bibnamefont{G\"otze}}, \bibinfo{journal}{J.
  Phys.: Condens. Matter} \textbf{\bibinfo{volume}{2}}, \bibinfo{pages}{8485}
  (\bibinfo{year}{1990}).

\bibitem[{\citenamefont{Gnann}(2009)}]{Gnann.2009}
\bibinfo{author}{\bibfnamefont{M.~V.} \bibnamefont{Gnann}},
  \emph{\bibinfo{title}{Analysis of schematic models of mode-coupling theory
  for colloids in external fields}}, \bibinfo{howpublished}{{D}iploma thesis,
  Universit{\"a}t Konstanz} (\bibinfo{year}{2009}).

\bibitem[{\citenamefont{Schnyder et~al.}(2011)\citenamefont{Schnyder,
  H\"of{}ling, Franosch, and Voigtmann}}]{Schnyder.2011}
\bibinfo{author}{\bibfnamefont{S.~K.} \bibnamefont{Schnyder}},
  \bibinfo{author}{\bibfnamefont{F.}~\bibnamefont{H\"of{}ling}},
  \bibinfo{author}{\bibfnamefont{T.}~\bibnamefont{Franosch}}, \bibnamefont{and}
  \bibinfo{author}{\bibfnamefont{{\relax Th}.}~\bibnamefont{Voigtmann}},
  \bibinfo{journal}{J. Phys.: Condens. Matt.} \textbf{\bibinfo{volume}{23}},
  \bibinfo{pages}{234121} (\bibinfo{year}{2011}).

\bibitem[{\citenamefont{Winter et~al.}(2012)\citenamefont{Winter, Horbach,
  Virnau, and Binder}}]{Winter.2012}
\bibinfo{author}{\bibfnamefont{D.}~\bibnamefont{Winter}},
  \bibinfo{author}{\bibfnamefont{J.}~\bibnamefont{Horbach}},
  \bibinfo{author}{\bibfnamefont{P.}~\bibnamefont{Virnau}}, \bibnamefont{and}
  \bibinfo{author}{\bibfnamefont{K.}~\bibnamefont{Binder}},
  \bibinfo{journal}{Phys. Rev. Lett.} \textbf{\bibinfo{volume}{108}},
  \bibinfo{pages}{028303} (\bibinfo{year}{2012}).

\bibitem[{\citenamefont{Khair and Brady}(2006)}]{Khair.2006}
\bibinfo{author}{\bibfnamefont{A.~S.} \bibnamefont{Khair}} \bibnamefont{and}
  \bibinfo{author}{\bibfnamefont{J.~F.} \bibnamefont{Brady}},
  \bibinfo{journal}{J. Fluid Mech.} \textbf{\bibinfo{volume}{557}},
  \bibinfo{pages}{73} (\bibinfo{year}{2006}).

\bibitem[{\citenamefont{Brader et~al.}(2009)\citenamefont{Brader, Voigtmann,
  Fuchs, Larson, and Cates}}]{Brader.2009}
\bibinfo{author}{\bibfnamefont{J.~M.} \bibnamefont{Brader}},
  \bibinfo{author}{\bibfnamefont{{\relax Th}.}~\bibnamefont{Voigtmann}},
  \bibinfo{author}{\bibfnamefont{M.}~\bibnamefont{Fuchs}},
  \bibinfo{author}{\bibfnamefont{R.~G.} \bibnamefont{Larson}},
  \bibnamefont{and} \bibinfo{author}{\bibfnamefont{M.~E.} \bibnamefont{Cates}},
  \bibinfo{journal}{Proc.\ Natl.\ Acad.\ Sci.\ USA}
  \textbf{\bibinfo{volume}{106}}, \bibinfo{pages}{15186}
  (\bibinfo{year}{2009}).

\bibitem[{\citenamefont{Fuchs and Mayr}(1999)}]{FuchsSE}
\bibinfo{author}{\bibfnamefont{M.}~\bibnamefont{Fuchs}} \bibnamefont{and}
  \bibinfo{author}{\bibfnamefont{M.~R.} \bibnamefont{Mayr}},
  \bibinfo{journal}{Phys. Rev. E} \textbf{\bibinfo{volume}{60}},
  \bibinfo{pages}{5742} (\bibinfo{year}{1999}).

\bibitem[{\citenamefont{Brillo et~al.}(2011)\citenamefont{Brillo, Pommrich, and
  Meyer}}]{MeyerSE}
\bibinfo{author}{\bibfnamefont{J.}~\bibnamefont{Brillo}},
  \bibinfo{author}{\bibfnamefont{A.~I.} \bibnamefont{Pommrich}},
  \bibnamefont{and} \bibinfo{author}{\bibfnamefont{A.}~\bibnamefont{Meyer}},
  \bibinfo{journal}{Phys. Rev. Lett.} \textbf{\bibinfo{volume}{107}},
  \bibinfo{pages}{165902} (\bibinfo{year}{2011}).

\end{thebibliography}
\bibliographystyle{apsrev}

\end{document}